\documentclass[a4paper,11pt]{article}
\pdfoutput=1
\usepackage{cite}
\usepackage{verbatim}
\usepackage{multirow}
\usepackage{jheppub}
\usepackage{hyperref,soul}
\usepackage{enumerate}
\usepackage{relsize}
\usepackage{mathrsfs}
\usepackage{pdflscape}
\usepackage{inputenc}
\usepackage{braket,xcolor,float}
\usepackage{bbold,amsmath,lipsum,amssymb,verbatim,cleveref}
\usepackage[T1]{fontenc}
\usepackage{graphicx,wrapfig}
\usepackage{jheppub}
\usepackage{bbm}
\usepackage{epsfig,dsfont}

\usepackage{bbm}
\usepackage{epsfig, amsmath, amssymb, dsfont, comment}
\usepackage{hyperref}

\hypersetup{
    colorlinks=true,
    linkcolor=blue,
    citecolor=blue,
    urlcolor=blue
}
\usepackage{graphicx,psfrag}
\usepackage{xcolor}
\usepackage{physics, cleveref}

\renewcommand{\L}{\mathcal{L}}
\newcommand{\nn}{\nonumber}
\newcommand{\be}{\begin{equation}}
\newcommand{\ee}{\end{equation}}
\newcommand{\ben}{\begin{equation}}
\newcommand{\een}{\end{equation}}
\newcommand{\bea}{\begin{eqnarray}}
\newcommand{\eea}{\end{eqnarray}}
\newcommand{\bA}{\begin{array}}
\newcommand{\eA}{\end{array}}
\newcommand{\bc}{\begin{center}}
\newcommand{\ec}{\end{center}}

\newcommand{\ra}{\rightarrow}
\newcommand{\del}{\partial}

\newcommand{\ie}{{\it i.e.}}
\newcommand{\eg}{{\it e.g.}}

\newcommand{\lan}{\langle}
\newcommand{\ran}{\rangle}

\newcommand{\scri}{{\cal I}}
\newcommand{\lra}{\leftrightarrow}

\numberwithin{equation}{section}

\allowdisplaybreaks[0]

\title{A Song of Spice and Ice: \\  \vspace{0.5 cm}
\Large{A diagonal dual to dS$_3$ \& its Galilean flat space limit}}
\author[a]{Arjun Bagchi,}
\emailAdd{abagchi} 
\author[a]{Sharang Rajesh Iyer,} 
\emailAdd{siyer23@iitk.ac.in}
\author[b]{and K. Narayan.}
\emailAdd{\ narayan@cmi.ac.in}\author{\\}
\affiliation[a]{Indian Institute of Technology Kanpur, Kanpur 208016, India.\\}
\affiliation[b]{Chennai Mathematical Institute,
H1 SIPCOT IT Park, Siruseri 603103, India.\\}

\abstract{We propose a diagonal CFT dual to asymptotically de Sitter spacetimes in three dimensions (3d), constructed from the holomorphic part of the CFT dual at the future boundary and the antiholomorphic part of the CFT dual at the past boundary, in part inspired by an antipodal map between the two CFT copies. This gives equal and opposite imaginary central charges for the holomorphic and antiholomorphic sectors. The corresponding diagonal holographic energy-momentum tensor expectation values are constructed from the Wavefunction at the future boundary and its conjugate. This enables a recovery of the entropy of $dS_3$, as well as a family of complex Kerr $dS_3$ spacetimes with two real horizons, by a Cardy formula.

We then explore the flatspace limit of our dS construction. With the central charges as above, the flatspace limit naturally induces a Galilean contraction on the dual diagonal 2d CFT, in contrast to the more familiar Carrollian limit when 3d Asymptotically Flat Spacetime (AFS) is approached from $AdS_3$. The corresponding limit for the complex Kerr family above leaves the inner horizon intact, resulting in flat space complex black holes ($\mathbb{C}$BH) that are closely related to flat space cosmologies which arise from BTZ black holes in AdS$_3$.  We recover the entropy of these $\mathbb{C}$BHs by the BMS-Cardy formula. 
}
\begin{document}

\maketitle

\section{Introduction}
The Holographic Principle has been our most trusted route to a theory of quantum gravity. Gravity in three spacetime dimensions is a very good laboratory for quantum gravity and holography. Einstein's theory of general relativity in three dimensions has a rather intriguing dichotomy. On the one hand, it has no propagating degrees of freedom and at first glance may appear trivial. On the other hand, it possesses very non-trivial solutions, the most famous of which are the Banados-Teitelboim-Zanelli (BTZ) black holes \cite{Banados:1992wn} in asymptotically 3d Anti de Sitter (AdS$_3$) spacetime. 

\medskip

Holography is of course best understood in AdS spacetimes through the AdS/CFT correspondence \cite{Maldacena:1997re,Gubser:1998bc,Witten:1998qj,Aharony:1999ti}. But the construction of holography beyond AdS is still very much a work in progress. Recently, there has been a cynosure of attention on aspects of flatspace holography following deep insights into infrared physics connecting asymptotic symmetries, soft theorems and memory effects \cite{Strominger:2017zoo}. The holographic constructions, mostly bottom up, have followed two complementary directions called Celestial \cite{Pasterski:2021raf, Raclariu:2021zjz} and Carrollian holography \cite{Bagchi:2025vri, Nguyen:2025zhg, Ruzziconi:2026bix}. One of the enduring themes has been attempts at constructing flatspace holography from a large radius limit of AdS/CFT, as we will review below.

\medskip

Ultimately, we wish to build a hologram for our universe, which has a small positive cosmological constant and is expected to evolve into a de Sitter universe at late times. dS also plays a pivotal role in the early inflationary phase of the universe. So a hologram of our universe is intimately tied to dS holography and $dS/CFT$
\cite{Strominger:2001pn,Witten:2001kn,Maldacena:2002vr,Anninos:2011ui}. 

\medskip

A cornerstone of holography is Strominger's recovery \cite{Strominger:1997eq} of the entropy of $AdS_3$ BTZ black holes
from the Cardy formula for the holographically dual 2d CFT.  It is fascinating to ask if de Sitter entropy \cite{Gibbons:1977mu} can be
realized as an asymptotic Cardy growth in holography. This is challenging since holography at the (natural) future boundary via $dS/CFT$
\cite{Strominger:2001pn,Witten:2001kn,Maldacena:2002vr,Anninos:2011ui} with the dictionary $Z_{CFT}=\Psi_{dS}$ \cite{Maldacena:2002vr} suggests exotic
nonunitary (ghost-like) dual Euclidean CFTs, which we do not really understand so far. In addition, for entirely Lorentzian de Sitter like spacetimes,
there are several indications that two copies of this dual CFT, at the future and past boundaries, are required to encode bulk observables
holographically. One might think the future and past CFT copies are related by bulk time evolution in some sense, but the dual CFTs are
entirely Euclidean with no time per se: we expect bulk time is emergent. An antipodal map used interestingly in \cite{Strominger:2001pn} 
could be regarded as a map between the two CFT copies.  Explorations of extremal surfaces and holographic entanglement (pseudo-)entropy
\cite{Narayan:2015vda,Narayan:2017xca,Doi:2022iyj,Narayan:2022afv}
give new insights into certain aspects, including the entropy, and are
still evolving:\ see also
\cite{Narayan:2017xca, Narayan:2019pjl, Arias:2019pzy, Cotler:2023xku, Fujiki:2026kca} for related discussions of de Sitter entropy as entanglement between two CFT copies.  Overall, the de Sitter hologram remains mysterious. See \eg\ \cite{Spradlin:2001pw,Anninos:2012qw,Galante:2023uyf} for reviews of various aspects.

\subsection*{Spice: The diagonal $dS_3$/CFT$_2$ hologram}

Our paper is about holography in three-dimensional bulk spacetimes. 

\medskip

In the first part of our paper, we propose a different construction of $dS_3$/CFT$_2$, which utilizes both boundaries and an antipodal matching to define a ``diagonal dual'' 2d CFT. The individual elements (viz. the antipodal map connecting two boundaries, the Wavefunction description) of course parallel various seminal constructions in the past \cite{Strominger:2001pn, Maldacena:2002vr}, but we believe our picture puts these separate ingredients together for the first time and presents a coherent new picture.   

\medskip

Let us briefly outline the construction for the impatient reader.
Asymptotically $dS_3$ spacetimes can be viewed as excitations about $dS_3$ encoded by the holographic energy-momentum tensor evaluated at the future and past boundaries $\scri^\pm$. Our dual diagonal CFT is defined from the two dual CFT copies at the future and past boundaries. It is constructed from the holomorphic part of the future CFT copy and the antiholomorphic part of the past CFT copy (sec.~\ref{sec:dS3-diagCFT2}){\footnote{Other constructions, e.g. taking the antiholomorphic part from the future boundary and the holomorphic part from the past boundary, lead to equivalent results.}}. The corresponding diagonal holographic energy-momentum tensor expectation values are constructed from the Wavefunction at $\scri^+$ and its conjugate. Thus, this diagonal CFT has imaginary central charge, but with opposite signs in its holomorphic and antiholomorphic sectors. The energy-momentum tensor can be seen to give complex weights $\ell_0, {\bar\ell}_0$ generically.

\medskip

In sec.~\ref{KdS3cmplx}, with our prescription, we recover the entropy of de Sitter space and of complex Kerr $dS_3$ spacetimes (with two real horizons). Through an interesting conspiracy of $\pm i$'s in the central charges and complex weights, the Cardy formula reveals a large real asymptotic growth matching the entropy of $dS_3$ and the complex Kerr $dS_3$ family.  

\medskip

There are several interesting technical
relations with analytic continuations from the $AdS_3$ BTZ case \cite{Strominger:1997eq}, but the overall picture of the bulk entropy is a consequence of two copies of the de Sitter CFT and thus not an analytic continuation per se. Our picture thus has some parallels with the studies in \cite{Balasubramanian:2001nb} (see also \cite{Park:1998qk,Klemm:2001ea,Balasubramanian:2002zh,Hao:2026xtk}) but is different overall.

\subsection*{Ice: from $dS_3$/CFT$_2$ to $Mink_3$/Galilean CFT$_2$}

The second part of our paper is on understanding the flat space limit of dS/CFT. 

\medskip

As we have said earlier, a popular route to flat space holography has been through AdS/CFT by looking at the infinite radius limit of AdS/CFT, following seminal works \cite{Polchinski:1999ry, Giddings:1999qu, Giddings:1999jq}. The recent advances include \cite{Bagchi:2023fbj, Bagchi:2023cen, Alday:2024yyj, Lipstein:2025jfj, Bagchi:2026bka}. The infinite radius limit of dS also lands one up in flatspace. It is thus interesting to ponder what lessons dS/CFT might have for flat holography, how it may be different from the AdS/CFT limit and ultimately how this triad of holographic dualities would fit in together. Taking initial steps towards understanding this holographic triangle, specifically in the case of three bulk dimensions, is one of our goals in this paper. 

\medskip

As we will go on to describe in sec.~\ref{Fire&Ice-0}, the flat space limit of AdS manifests itself as a Carrollian contraction of the dual CFT where the speed of light ($c$) in the field theory goes to zero, leading to a change in the underlying algebra to match what happens in the bulk \cite{Bagchi:2010zz, Bagchi:2012cy,Duval:2014uva}. In 2d theories, there is an interesting interplay between the Carrollian $(c\to0)$ and Galilean $(c\to \infty)$ limits, which leads to isomorphic symmetry algebras. 
The 2d Carrollian and 2d Galilean CFTs are thus governed by the same algebra which is also isomorphic to the asymptotic symmetry algebra of 3d Asymptotically Flat Spacetimes (AFS), called the 3d Bondi-van der Burgh-Metzner-Sachs (BMS$_3$) algebra \cite{Barnich:2006av, Bagchi:2010zz, Duval:2014uoa}. 

\medskip

We will show that our construction of dS$_3$/CFT$_2$ naturally leads us to the Galilean contraction of the diagonal relativistic 2d CFT when the flatspace limit is implemented in the bulk, in contrast to the Carrollian contraction that arises in the AdS case. 

\medskip

We will go on to consider the flat limit of the complex Kerr $dS_3$
black holes (sec.~\ref{sec:CBH}). It is known that 3d AFS does not admit any black hole solutions \cite{Ida:2000jh}. However, our flatspace limit will lead to black holes and we will evade the theorem above since these novel black holes have complex metrics. These Complex Black Holes ($\mathbb{C}$BH) are analytic continuations of Flat Space Cosmologies (FSC) \cite{Cornalba:2002fi, Bagchi:2012xr} that one obtains as shifted boost orbifolds of Minkowski spacetimes. $\mathbb{C}$BHs have horizons and an associated entropy, which we show can be reproduced from the BMS-Cardy formula \cite{Bagchi:2012xr}, albeit with unfamiliar weights in the Galilean CFT.

\section{\texorpdfstring{$dS_3$}{dS3} holography, a diagonal \texorpdfstring{$CFT_2$}{CFT2} and antipodal map}
\label{sec:dS3-diagCFT2}

$dS/CFT$ dualities \cite{Strominger:2001pn,Witten:2001kn,Maldacena:2002vr,Anninos:2011ui},
argue for hypothetical dual Euclidean CFT holographic descriptions at the spacelike future (or past)
boundary of de Sitter space, with dictionary \cite{Maldacena:2002vr}
\be\label{Zcft=Psids}
Z_{CFT}=\Psi_{dS} \sim e^{iI_{cl}}\,,\quad \lan O_k O_{k'}\ran
= \frac{\delta^2\log Z_{CFT}}{\delta\varphi_k^0\delta\varphi_{k'}^0}\,,\quad
\lan\varphi_k\varphi_{k'}\ran \sim
\int D\varphi\,\varphi_k\,\varphi_{k'}\,|\Psi_{dS}[\varphi_k]|^2\,,
\ee
where $Z_{CFT}$ is the CFT partition function, $\Psi_{dS}$ is the Hartle-Hawking
Wavefunction of the Universe \cite{Hartle:1983ai}, and 
$\varphi$ are bulk modes with future boundary values $\varphi^0$
that couple to the dual operators $O$.  This suggests the duals are
nonunitary, ghost-like, CFTs with negative or imaginary central
charges, as we will see in detail below. A central point to note above
is that while dual operator correlation functions require a single
copy of the Wavefunction $\Psi_{dS}=Z_{CFT}$ above, bulk observables
require two copies, $\Psi_{dS}$ and $\Psi_{dS}^*$, taken suggestively
to arise from the future boundary $\scri^+$ and the past boundary
$\scri^-$. The case of $dS_3$,
which has imaginary central charge, is particularly intriguing.  In what follows, we will describe
a construction of a diagonal CFT from the holomorphic and
antiholomorphic parts of the future and past CFT copies as the dual to
Lorentzian asymptotically $dS_3$ spacetimes. This allows a recovery of
the entropy of $dS_3$ and the family of complex Kerr $dS_3$ spacetimes 
from the Cardy formula for the diagonal CFT.

\medskip

The imaginary central charge of $dS_3$ can
be noted from the calculation of the dual 2-point correlation function
of operators dual to bulk scalar modes representing analogs of
appropriate graviton modes from (\ref{Zcft=Psids}).
For $dS_3$, the semiclassical Wavefunction for the dual stress tensor 2-point function above gives imaginary central charge\
$c_{dS_3}\sim i\frac{l}{G_3}$ at the future boundary (more on this
later). This can also be seen via other probes. Analogs of the Ryu-Takayanagi formulation in dS reveal bulk 
extremal surfaces anchored at $\scri^+$ 
\cite{Narayan:2015vda,Narayan:2017xca,Doi:2022iyj,Narayan:2022afv,
Narayan:2023zen,Doi:2023zaf,Nanda:2025tid,Narayan:2026wzp}
containing complex or timelike components, with complex areas, 
best regarded as pseudo-entropies \cite{Nakata:2020luh}. For instance we obtain\ 
$S = i\frac{l}{2G_3}\log \frac{R_c}{l}+ \frac{\pi\,l}{4G_3}$\ for
maximal subregions in no-boundary $dS_3$.\
The imaginary (timelike) part from the Lorentzian region is of 
Calabrese-Cardy form ${\frac{c}{3}}\log\frac{l}{\epsilon}$
with $R_c\equiv \frac{l^2}{\epsilon}$ the cutoff at ${\scri}^+$ and
$c=i\frac{3l}{2G_3}$ the central charge. The second, real term above
from the Euclidean hemisphere is half $dS_3$ entropy. Other recent 
discussions of $dS_3/CFT_2$ include \eg\ \cite{Hikida:2022ltr}.

\subsection*{$dS_3$ and the holographic energy-momentum tensor}

From this $dS/CFT$ point of view, asymptotically $dS_3$ spacetimes are perhaps best viewed as excitations over pure $dS_3$, characterized by the corresponding holographic energy momentum tensor expectation values (in line with \cite{Das:2013mfa}).  Conversely, we view generic asymptotically $dS_3$ geometries as gravitational saddles dual to $CFT_2$ states/ensembles with specified stress tensor expectation values. To quantify this, we now evaluate the holographic stress tensor. The regulated 3d action with counterterms at $\scri^+$ (taken at $r\ra\infty$) is
\be\label{action-Ir}
I_r = \frac{1}{16\pi G_3} \int d^3x \sqrt{-g} \left(R-\frac{2}{l^2}\right)
- \frac{2}{16\pi G_3}\int d^2x \sqrt{h} K
+ \frac{2}{16\pi G_3} \int d^2x \frac{\sqrt{h}}{l}\,,
\ee
Let us write this asymptotically $dS_3$ metric in the general form
($\{i,j\}=\{z,{\bar z}\}$)
\be
ds^2 = -N_r^2dr^2 + h_{ij}dx^i dx^j\,,\qquad n=N_rdr\,,
\qquad
K_{ij} = \Gamma_{ij}^rn_r=-\frac{1}{2 N_r} \del_rh_{ij}\,,
\ee
(with the future boundary at $r\ra\infty$, the outward pointing normal
is $n\propto +dr$;\ and $g_{r i}=0$ for our cases, so $g^{rr}=g_{rr}^{-1}$
and $\Gamma^r_{ij}$ has only the term above).\
Then, using the $dS/CFT$ dictionary \cite{Maldacena:2002vr} with the semiclassical
Wavefunction $\Psi_{dS}\sim e^{iI_r}$, the boundary stress tensor at $\scri^+$ is
\begin{equation}\label{Tij-D}
    T_{ij}^+ = \frac{2}{\sqrt{{\hat h}}}\frac{\delta \log Z_{CFT}}{\delta\hat h^{ij}} = \frac{2}{\sqrt{{\hat h}}}\frac{\delta \log\Psi_{dS}}{\delta\hat h^{ij}}
= i \frac{1}{8\pi G} (K_{ij}-Kh_{ij}-\frac{1}{l}h_{ij})\,,
\end{equation}
evaluated via the regulated action $I_r$ in (\ref{action-Ir}) above.
(For general dimension, rescaling by the $\epsilon$-factors in
$\sqrt{h}h^{ij}$ to match the boundary value
$\sqrt{{\hat h}}{\hat h}^{ij}$ is more nontrivial)
This then gives
\be\label{TijdS3gen}
T_{ij}^+ = {i\over 8\pi G_3} \Big( -{1\over 2 N_r} \del_rh_{ij} +
{1\over 2 N_r} h_{ij}\ (h^{kl}\del_rh_{kl})
- {1\over l} h_{ij} \Big) .
\ee
Pure $dS_3$ in the Poincar\`e slicing with only nonzero $h_{z{\bar z}}$
gives vanishing $T_{zz}^+, T_{{\bar z}{\bar z}}^+$, as we will elaborate
later; general asymptotically $dS_3$ metrics will contain various
nonzero $L_n, {\bar L_n}$ modes.

\subsection*{The future and past CFTs}

A simple way to see the $dS_3$ central charge proceeds analogous to
\cite{Balasubramanian:1999re} for $AdS_3$.
Boundary conformal transformations\
$z\ra z+w,\ \ {\bar z}\ra {\bar z}+{\bar w}$\ induce a Weyl factor on
$r$ in the bulk metric and thereby extra cross-terms, departing from
the asymptotic metric. So consider bulk diffeomorphisms that act in a near-boundary expansion as boundary conformal transformations undone by a Weyl transformation, thus preserving the asymptotic metric to
leading order. Along the lines of the Brown-Henneaux $AdS_3$ analysis,
the asymptotic Poincar\`e $dS_3$ metric and allowed diffeomorphisms
(given by arbitrary $w(z), {\bar w}({\bar z})$) give
\bea\label{dS3defm-wwbar}
&& z\ra z+w+{l^2\over 2r^2} {\bar\del}^2{\bar w}\,,\ \ \
{\bar z}\ra {\bar z}+{\bar w}+{l^2\over 2r^2} \del^2w\,,\ \ \
r\ra r-{r\over 2} (\del w+{\bar\del}{\bar w})\,, \nn\\
&& ds^2 = -{l^2\over r^2} dr^2 + {r^2\over l^2} dzd{\bar z}\ \ra\
-{l^2\over r^2} dr^2 + {r^2\over l^2} dzd{\bar z} +
{l^2\over 2} \del^3w\, dz^2
+ {l^2\over 2} {\bar\del}^3{\bar w}\, d{\bar z}^2 \,. \qquad
\eea
Then
\bea\label{dS3CFT2:Tzz-c}
\delta T_{zz}^+ = {i\over 8\pi G}\,\Big({1\over l} \delta h_{zz}\Big)
= {i l\over 16\pi G} \del^3w\  \equiv\  {c\over 24\pi} \del^3 w 
\ \ \ &\Rightarrow & \ \ \ c= i{3 l\over 2G}\,,\nn\\ [1mm]
\delta T_{{\bar z} {\bar z}}^+ = {i\over 8\pi G}\,\Big({1\over l}
\delta h_{{\bar z} {\bar z}}\Big)
= {i l\over 16\pi G} \del^3{\bar w}\  \equiv\ 
{{\bar c}\over 24\pi} {\bar\del}^3 {\bar w} 
\ \ \ &\Rightarrow & \ \ \ {\bar c} = i{3 l\over 2G}\,,
\eea
where the first expression for $\delta T^+$ (in each line) is from the holographic stress tensor
(\ref{TijdS3gen}) evaluated on the above deformed metric 
(\ref{dS3defm-wwbar}), and the second expression
arises from the expected Schwarzian contribution to the CFT stress tensor for an infinitesimal 2d conformal transformation $z\ra z+w(z)$. Overall, this is essentially redoing the calculation in \cite{Strominger:2001pn}
and \cite{Balasubramanian:2001nb}, but with the Wavefunction now
which then gives the extra $i$. This can be seen to be tantamount
to analytic continuation from the $AdS_3$ central charge calculation
in \cite{Balasubramanian:1999re}.

\medskip

It is important to note that both $T_{zz}^+$ and $T_{{\bar z} {\bar z}}^+$
are evaluated via the Wavefunction $\Psi_{dS}$ at the future boundary
${\scri}^+$. A similar calculation at the past boundary $\scri^-$ gives
\be
T_{ij}^-
   = {2\over\sqrt{{\hat h}}}{\delta \log Z_{CFT}^*\over\delta
  {\hat h}^{ij}} = {2\over\sqrt{{\hat h}}}{\delta \log\Psi_{dS}^*\over\delta
  {\hat h}^{ij}}
 = -i\,{2\over\sqrt{{\hat h}}}{\delta \log I_{r}\over\delta
   {\hat h}^{ij}}\,,
 \ee
where we have taken the Wavefunction to satisfy\
$\Psi_{dS}[{\cal I}^-]=\Psi_{dS}[{\cal I}^+]^*$.\ This is equivalent 
to evaluating the regulated bulk action $I_r$ in (\ref{action-Ir}) at
the past boundary $\scri^-$, with its appropriate coordinate ranges.
Thus finally $\Psi_{dS}[{\cal I}^-]\sim e^{-iI_r}$
with $I_r=I_r^+$ in (\ref{action-Ir}).


\subsection*{A diagonal CFT}

Thus far, we have two dual CFT copies, $CFT_F\times CFT_P$, at the
future and past boundaries $\scri^\pm$, with
the energy momentum tensor sectors satisfying $T^\pm_{z{\bar z}}=0$ and
\be
Z_{CFT}^+[\scri^+]:\quad T_{zz}^+,\ {\bar T}_{{\bar z} {\bar z}}^+\ ;\qquad
Z_{CFT}^-[\scri^-]:\quad T_{zz}^-,\ {\bar T}_{{\bar z} {\bar z}}^-\,.
\ee
We now define a diagonal CFT from the holomorphic part of the
future CFT copy and the antiholomorphic part of the past CFT copy. So
for this diagonal CFT, we have $T^d_{z{\bar z}}=0$ and
\bea\label{TzzPsi-TzbzbPsi*}
&&   {T}^d_{zz} = T_{zz}^+ 
   = {2\over\sqrt{{\hat h}}}{\delta \log Z_{CFT}\over\delta
  {\hat h}^{zz}} = {2\over\sqrt{{\hat h}}}{\delta \log\Psi_{dS}\over\delta
  {\hat h}^{zz}}
 = i\,{2\over\sqrt{{\hat h}}}{\delta \log I_{r}\over\delta
   {\hat h}^{zz}}\,, 
 \nn\\ [1mm]
&& {\bar T}^d_{{\bar z} {\bar z}} = {\bar T}_{{\bar z} {\bar z}}^-
   = {2\over\sqrt{{\hat h}}}{\delta \log Z_{CFT}^*\over\delta
  {\hat h}^{{\bar z} {\bar z}}} = {2\over\sqrt{{\hat h}}}{\delta \log\Psi_{dS}^*\over\delta
  {\hat h}^{{\bar z} {\bar z}}}
 = -i\,{2\over\sqrt{{\hat h}}}{\delta \log I_{r}\over\delta
   {\hat h}^{{\bar z} {\bar z}}} = -{\bar T}_{{\bar z} {\bar z}}^+\,,\qquad
\eea
With this definition, we have an overall $-i$ instead of $i$ in
(\ref{TijdS3gen}) for ${\bar T}^d_{{\bar z}{\bar z}}$. This gives
\bea\label{dS3CFT2:Tzz-c,Tzbzb-cbar-1}
\delta T^d_{zz} = {i\over 8\pi G}\,\Big({1\over l} \delta h_{zz}\Big)
= {i l\over 16\pi G} \del^3w\ \ \equiv\ \ {c\over 24\pi} \del^3 w
&\Rightarrow & c= i{3 l\over 2G}\,,\nn\\ [1mm]
\delta {\bar T}^d_{{\bar z} {\bar z}} = -{i\over 8\pi G}\,\Big({1\over l}
\delta h_{{\bar z} {\bar z}}\Big)
= -{i l\over 16\pi G} \del^3{\bar w}\ \ \equiv\ \
{{\bar c}\over 24\pi} {\bar\del}^3 {\bar w}
&\Rightarrow & {\bar c} = -i{3 l\over 2G}\,.
\eea
It is important to note that in the last line, the first factor
$-{i l\over 16\pi G}$ arises from the holographic calculation via
$\Psi_{dS}^*$ while the second term factor ${{\bar c}\over 24\pi}$
is from the standard Schwarzian for the antiholomorphic part of the
transformation of the stress tensor under a conformal transformation
in the diagonal CFT. The extra minus sign then leads to ${\bar c}$ above.\\
The diagonal CFT is perhaps best thought of as living in some abstract space, a hybrid of the future and past boundaries.


\subsection*{An antipodal map}
It is striking to note that the above hybrid construction of this diagonal CFT is equivalent to imposing an antipodal map on the two CFT copies. To elaborate, with the $CFT$s defined on the plane, restricting to conformal deformations implies that $T_{z{\bar z}}^\pm=0$, which can be checked from (\ref{TijdS3gen}). In Lorentzian de Sitter, the past and future boundary slices are not independent, but related by bulk time evolution. The dual CFTs are Euclidean with no time per se: the two CFT copies are instead related by an antipodal map \cite{Strominger:2001pn}
\be\label{antipodal}
z^{(-)}=-{1\over{\bar z}^{(+)}}\,,\qquad {\bar z}^{(-)}=-{1\over z^{(+)}}\,.
\ee
This is equivalent to identifying antipodal points of the sphere $S^2$
at $\scri^+$ with those of the $S^2$ at $\scri^-$: specifically,
$\theta^{(-)}\lra \pi-\theta^{(+)},\ \phi^{(-)}\lra \phi^{(+)}+\pi$. In
terms of the cylinder representation of the boundary (in the static
coordinatization), we have\
$t^{(-)}\lra -t^{(+)},\ \phi^{(-)}\lra \phi^{(+)}+\pi$.\
Since the antipodal map (\ref{antipodal}) relates the holomorphic and antiholomorphic coordinates, the diagonal CFT stress tensor has its holomorphic part defined naturally from the holomorphic part of the future CFT copy, and its antiholomorphic part is defined from the holomorphic part of the past CFT copy. Specifically \footnote{We were assisted in this calculation initially by Claude (Opus 5.5).}
\be
T^d(z) = T^+_{zz}(z)\,,\qquad T^d({\bar z}) = T^-_{zz}(z^{(-)}) = 
{1\over {\bar z}^4}T_{zz}^-\Big(-{1\over {\bar z}}\Big)\,.
\ee
This has some parallels with the picture in \cite{Narayan:2017xca} of
pure de Sitter space as dual to two copies of the dual CFT in a thermofield double type future-past (time-)entangled state (see also related discussions in \cite{Narayan:2019pjl,Arias:2019pzy,Cotler:2023xku,Fujiki:2026kca}): the details are quite different however.

The antipodal map (\ref{antipodal}) can be seen to show consistency of
these central charges: for instance, defining $w_i=-{1\over {\bar z}_i}$
gives\ $w_{12}={{\bar z}_{12}\over {\bar z}_1 {\bar z}_2}$ so
\be
\lan T^d({\bar z}_1) T^d({\bar z}_2) \ran = {1\over {\bar z}_1^4 {\bar z}_2^4}
\lan T^-(w_1)T^-(w_2) \ran \sim
{1\over {\bar z}_1^4 {\bar z}_2^4} \ {{\bar c}/2\over w_{12}^4}
= {{\bar c}/2 \over {\bar z}_{12}^4}\,.
\ee
The rest of the antiholomorphic part of the diagonal stress tensor OPE can also be shown to be consistently defined by that of $T^-$ while\ $\lan T^d(z_1)T^d(z_2)\ran \sim {c/2\over z_{12}^4}$\ as usual for the holomorphic sector.

In what follows, we will use the holographic stress tensor (\ref{TzzPsi-TzbzbPsi*}) for the diagonal CFT encoding of our bulk geometries. We now define $z={t\over l}-i\phi,\ {\bar z}={t\over l}+i\phi$,
and
\be\label{TzzTzbzb-2}
l^2T_{tt} - T_{\phi\phi} + 2il T_{t\phi} = T_{zz}\,, \qquad
l^2T_{tt} - T_{\phi\phi} - 2il T_{t\phi} = {\bar T}_{{\bar z}{\bar z}}\,,\qquad
T_{z{\bar z}}\propto l^2T_{tt} + T_{\phi\phi}\,.
\ee
For $dS$ deformations preserving conformal invariance, we have
$T_{z{\bar z}}=0$ so $T_{\phi\phi}=-l^2T_{tt}$.\ So this allows a
simple reconstruction of\ $T_{tt},\ T_{\phi\phi},\ T_{t\phi}$,\ from
$T_{zz},\ {\bar T}_{{\bar z}{\bar z}}$,\ consistent with the prescription
(\ref{TzzPsi-TzbzbPsi*}). 
With this definition for the antiholomorphic part with the minus sign, all the antiholomorphic Virasoro modes will also acquire a minus sign.
For more general deformations of $dS_3$, \ie\ more general cosmologies, we expect that $T_{z{\bar z}}\neq 0$: the deformations will not preserve any conformal invariance. It is unclear if the above diagonal prescription (\ref{TzzPsi-TzbzbPsi*}) can be consistently defined then, since the ``top-bottom'' time symmetry underlying this prescription (as well as the discussions in \cite{Narayan:2017xca}) will be broken.

\section{Complex Kerr \texorpdfstring{$dS_3$}{dS3} entropy: Cardyology}
\label{KdS3cmplx}

As a concrete example to illustrate our diagonal $CFT_2$ dual to Lorentzian asymptotically $dS_3$ spaces proposed in the previous section, we consider the family of complex Kerr $dS_3$ spacetimes (which also contains $dS_3$) and their dual holographic energy-momentum tensor values. This allows us to recover their entropy by the Cardy formula of the diagonal CFT.

\medskip

Consider the asymptotically $dS_3$ spacetime with real parameters
$E, j$, and metric 
\bea\label{KdS3:imagJ}
&&\quad ds^2 = {(r^2-r_+^2)(r^2-r_-^2)\over r^2 l^2} dt^2 -
{l^2 r^2\over (r^2-r_+^2)(r^2-r_-^2)} dr^2 +
r^2 \Big(d\phi+i{r_+r_-\over r^2 l} dt\Big)^2 \nn\\ [1mm]
&&\qquad\ \ = -\Big( 1-8GE -{r^2\over l^2} \Big) dt^2 +
{dr^2\over 1-8GE -{r^2\over l^2} - {16 G^2j^2\over r^2}} + r^2 d\phi^2
+ i\,8Gj dt d\phi\ ,\nn\\
&& \qquad\qquad {r_+^2+r_-^2\over l^2} = 1-8GE \,,\qquad
{r_+r_-\over 4G l} \equiv j\,.
\eea
We have expressed it in various equivalent forms, useful for our purposes below.\ (There are close relations to analytic continuations from the $AdS_3$ BTZ black hole: see sec.~\ref{sec:BTZcontns} later.)\ These are solutions to Einstein gravity with positive cosmological constant, satisfying\
$R_{\mu\nu}={2\over l^2} g_{\mu\nu}$\,,\ with\ $R={6\over l^2}$.

\medskip

For $j=0$, the complex metric (\ref{KdS3:imagJ}) collapses to Schwarzschild $dS_3$\
\be\label{SdS3:J=0}
ds^2 = -\Big( 1-8GE -{r^2\over l^2} \Big) dt^2 +
{dr^2\over 1-8GE -{r^2\over l^2}} + r^2 d\phi^2\,,\qquad
0<8GE<1\,.
\ee
As is familiar from \eg\ \cite{Spradlin:2001pw}, the vicinity of $r=0$ has the spatial metric\ $d\sigma^2={dr^2\over 1-8GE}+r^2d\phi^2$\,,\ which exhibits a conical singularity with deficit angle $2\pi-2\pi\sqrt{1-8GE}$, giving the condition on $E$ above: at $8GE=1$ the space closes up completely. This is unlike higher dimensional Schwarzschild de Sitter black holes, which acquire genuine curvature singularities at $r=0$. \ 
For $E=0$, the metric (\ref{SdS3:J=0}) becomes pure $dS_3$ in the static coordinatization and $r\leq l$ is the static patch where $t$ is a time coordinate. The regions $r>l$ describe the future and past (Milne) universe patches.

\medskip

The asymptotic structure of the metric (\ref{KdS3:imagJ}) for any $j$ is 
\be
r\gg l:\qquad ds^2 = {r^2\over l^2} dt^2
- {l^2 dr^2\over r^2} + r^2 d\phi^2 + 8iGj dt d\phi\,,
\ee
of the general $dS_3$ form (with rotation) near the future boundary $\scri^+\equiv (t,\phi)$\ [$r\equiv$ time].

In the static patch $r_-<r<r_+$, the $t$-coordinate is timelike and the metric (\ref{KdS3:imagJ}) can be Euclideanized by Wick-rotating $t$: for pure $dS_3$ this gives a regular 3-sphere. In the present case, we obtain a real Euclidean metric describing a squashed twisted 3-sphere (see App.~\ref{KdS3Euclst}).

\subsection{Outer and inner horizon entropies}\label{cmplxKdS3ent}

Although we have a complex metric (\ref{KdS3:imagJ}), the blackening factor has real $r_\pm$, \ie\ two real horizons, located at
\bea\label{KdS3r+r-}
&& r_-={4Gjl\over r_+}\ \ra\ (1-8GE)l^2r_+^2=r_+^4+16G^2j^2l^2\,,\nn\\
\ra &&
r_\pm^2={1\over 2}(1-8GE)l^2\pm {1\over 2}\sqrt{(1-8GE)^2l^4-64G^2j^2l^2}\,.
\eea
With $r_->0$ we have $j>0$.\ This gives the outer and inner horizon areas 
\be\label{KdS3entr+}
S={2\pi r_\pm\over 4G} = {\pi r_\pm\over 2G} = real\,,\qquad\qquad
64G^2j^2l^2 \leq (1-8GE)^2l^4\,,
\ee
which we regard as real entropies of both horizons. This reality stems from the parameter $j$ being real.

\medskip

With $j=0$, this is\ the Schwarzschild $dS_3$ entropy\
$S={\pi l\over 2G}\sqrt{1-8GE}$\,, which reduces to $dS_3$ entropy for $E=0$. This is the maximal entropy in asymptotically $dS_3$ spaces: the black hole with $E$ or $j$ nonzero reduces the entropy. For $8GE=1, j=0$, the space closes off entirely, and the entropy vanishes. For $j={(1-8GE) l\over 8G}$, we hit extremality with $r_+=r_-$: this is a Nariai-like limit of (\ref{KdS3:imagJ}) with $dS_2\times S^1$, but with a complex fibration.

\subsection{\texorpdfstring{$T_{ij}$}{Stress tensor} and the dual CFT weights}\label{sec:holoTij}

Viewing the complex Kerr $dS_3$ spaces (\ref{KdS3:imagJ}) as excitations over pure $dS_3$ with planar (Poincar\`e) slicing, we would like to understand the dual diagonal $CFT_2$ interpretation and the entropy. Towards this, first define dimensionless complex coordinates as
\be
z={t\over l}-i\phi\,,\qquad {\bar z}={t\over l}+i\phi\,.
\ee
Then the complex metric (\ref{KdS3:imagJ}) is recast as
\bea\label{KdS3:imagJ-zzbar}
&& ds^2 = -{dr^2\over {r^2\over l^2}-1 + 8GE + {16 G^2j^2\over r^2}}
+ \Big( r^2 - {(1-8GE) l^2\over 2} \Big) dz d{\bar z} \nn\\ [1mm]
&&\qquad\qquad + \Big( - {(1-8GE) l^2\over 4} - 2Gj l \Big) dz^2
+ \Big( - {(1-8GE) l^2\over 4} + 2Gj l \Big) d{\bar z}^2\ .\quad
\eea
For $E=0, j=0$, this is pure $dS_3$ in the static coordinatization: relative to the planar slicing, there are subleading components which play important roles as we will see below. We now evaluate the dual holographic energy-momentum tensor for the diagonal CFT defined via (\ref{TzzPsi-TzbzbPsi*}) using (\ref{TijdS3gen}), (\ref{Tij-D}), for each component. Since we are using dimensionless $z, {\bar z}$, there are overall $l^2$-factor differences that we must append to match those. The simplest case is pure $dS_3$ in the Poincar\`e slicing with metric\ $ds^2=-{l^2\over r^2} dr^2 + {r^2\over l^2} dzd{\bar z}$\,:\ this has only nonzero $h_{z{\bar z}}$, and it is straightforward to see that
\be
T_{zz}^d,\ {\bar T}^d_{{\bar z}{\bar z}}\ \propto\ \pm {i\over 8\pi G_3}
\left(-{r^2\over l^3}
+{r\over 2l}{r^2\over l^2}\Big(2{l^2\over r^2}{2r\over l^2}\Big)
-{r^2\over l^3}\right) = 0\,,
\ee
so planar $dS_3$ is like a vacuum: global $dS_3$ is dramatically different as we will see soon. In evaluating (\ref{TijdS3gen}) for the complex Kerr $dS_3$ metric in the form (\ref{KdS3:imagJ-zzbar}), note that $h_{zz}$ and $h_{{\bar z}{\bar z}}$ are constant: we expand the second term ${1\over 2 N_r} (h^{kl}\del_rh_{kl})$ at large $r$ and simplify. This finally gives the dimensionless expectation values $\ell_0, {\bar\ell_0}$,
\begin{subequations}\label{KdS3L0L0bar}
  \bea
T^d_{zz} \sim
{i\over 8\pi G} \Big( - {(1-8GE) l\over 4} - 2Gj \Big)
\equiv {1\over 2\pi}\,\ell_0\,, \\
{\bar T}^d_{{\bar z}{\bar z}} \sim
-{i\over 8\pi G} \Big( - {(1-8GE) l\over 4} + 2Gj \Big)
\equiv {1\over 2\pi}\,{\bar \ell_0}\,,
\eea  
\end{subequations}
semiclassically from the gravity side. This is equivalent to 
defining\ $\ell_0=\int d\phi\,T^d_{zz}$\ and\ 
${\bar\ell}_0=\int d\phi\,{\bar T}^d_{{\bar z}{\bar z}}$\,. The 
Euclidean CFT lives on the cylinder with coordinates 
$z,{\bar z}\equiv t,\phi$ and the weights $\ell_o,\ {\bar\ell}_0$
are evaluated on a $t=const$ slice integrating over the $\phi$-circle, 
which gives the $2\pi$-factor above.\ There are parallels with defining
the asymptotic bulk gravitational charges (mass and angular momentum) as
in \cite{Balasubramanian:2001nb} at $\scri^+$ (on a $t=const$ slice):  
that however uses the quasilocal stress tensor defined by the action, 
rather than the Wavefunction, thus differing by $i$-factors.

\medskip

Likewise, it can be checked that both
$T^+_{z{\bar z}}$ and $T^-_{z{\bar z}}$ vanish, upon expanding the terms
in (\ref{TijdS3gen}) at large $r$. Thus, along with the central charges, the dual diagonal CFT data encoding the complex Kerr $dS_3$ space is
\begin{subequations}\label{cbar=-c,l0bar-}
    \bea
{1\over 2}\Big( -{il\over 8 G}\,(1-8GE) - ij \Big) \equiv \ell_0\,,\ \ &&\ \
-{1\over 2}\Big( -{il\over 8 G}\,(1-8GE) + ij \Big) \equiv
{\bar \ell_0}\,, \\ [1mm]
c= i{3 l\over 2G}\,,\ \ &&\ \ {\bar c} = -i{3 l\over 2G}\,.
\eea
\end{subequations}
The weights $\ell_0, {\bar\ell_0}$, are pure imaginary. (Note in
particular that global $dS_3$, which is (\ref{KdS3:imagJ-zzbar}) with
$E=0, j=0$, has nontrivial weights.)
This suggests that the corresponding dual CFT operators $L_0, {\bar L_0}$
satisfy
\be\label{L0L0barConjugations}
\ell_0^*=-\ell_0\,,\ \ \ {\bar\ell_0}^*=-{\bar\ell_0}\qquad
\Rightarrow\qquad L_0^\dag = -L_0\,,\ \ \ {\bar L_0}^\dag = -{\bar L_0}\,.
\ee
Thus $H\equiv L_0+{\bar L_0}$ is anti-Hermitian, satisfying
$H^\dag=-H$. The complex weights, the corresponding operators and their conjugations are unusual from the point of view of ordinary unitary 2d CFTs.
Note that $L_0^\dag=-{\bar L_0}$ is also a valid conjugation if there
is no angular momentum (\ie\ $j=0$). The spin part alone satisfies
$L_0^\dag={\bar L_0}$, \eg\ with $8GE=1$ where
$L_0=-{ij\over 2}\,,\ {\bar L_0}={ij\over 2}$\,.\ 
It is worth noting that the conjugations above are distinct
from the situation for the real metric (\ref{KdS3:realJ}) in
\cite{Balasubramanian:2001nb} which gives\ $L_0^\dag=-{\bar L_0}$
using (\ref{KdS3L0L0bar}) with imaginary $j=-iJ$. Note also that
the conjugations above are distinct from those in
\cite{Doi:2024nty,Hao:2025btl} as well.

\subsection{The entropies via Cardy}\label{KdS3Cardy}

The Cardy formula for the asymptotic growth of states in a 2d CFT is
\be\label{cardy}
S = 2\pi \sqrt{{c \ell_0\over 6}} + 2\pi \sqrt{{{\bar c} {\bar\ell_0}\over 6}}
\ee
with the left and right central charges $c, {\bar c}$ and weights
$\ell_0, {\bar\ell_0}$.\  Using (\ref{cbar=-c,l0bar-}), we obtain
\bea\label{KdS3entr+2}
S_{r_+} &=& \pi \sqrt{{il\over 2G} \Big(-il {1-8GE\over 8G}-ij\Big)}
+ \pi \sqrt{-{il\over 2G} \Big(il{1-8GE\over 8G}-ij\Big)} \nn\\
&=& {\pi\over 2G} \left({1\over 2}\sqrt{(1-8GE)l^2+8Gjl}
+ {1\over 2}\sqrt{(1-8GE)l^2-8Gjl} \right) = {\pi r_+\over 2G}\ .
\eea
The last match is obtained by comparing the bracket-squared with
$r_+^2$ in (\ref{KdS3r+r-}).\
Thus, we recover the outer horizon entropy via Cardy. For pure $dS_3$,
we have $E=0, j=0$, in (\ref{KdS3:imagJ-zzbar}): this is
\be\label{dS3:imagJ-zzbar}
ds^2 = -{dr^2\over {r^2\over l^2}-1}
+ \Big( r^2 - {l^2\over 2} \Big) dz d{\bar z} 
+ \Big( - {l^2\over 4} \Big) dz^2
+ \Big( - {l^2\over 4} \Big) d{\bar z}^2\ .
\ee
This is sensitive to the underlying sphere foliation, which is
encoded by the subleading terms at large $r$ relative to Poincar\`e
$dS_3$. The complex weights (\ref{cbar=-c,l0bar-}) are thus nontrivial
and the Cardy formula (\ref{cardy}) for pure $dS_3$ gives
\be\label{cardydS3ent}
S_{l} = \pi \sqrt{{il\over 2G} \Big({-il\over 8G}\Big)}
+ \pi \sqrt{-{il\over 2G} \Big({il\over 8G}\Big)}
= {\pi l\over 2G}\,,
\ee
which is precisely de Sitter entropy in 3d. The large real
growth arises through an interesting conspiracy of $\pm i$'s in the
central charge and the complex weights.

\medskip

The exotic structures above are not surprising from the point of
view of the $dS_3$ dual being an exotic ghost-like CFT with imaginary
central charge and complex conformal dimensions.
For such a nonunitary CFT to have large real asymptotic growth of
states, the contributing excitation sectors must have large
$\mp {\rm Im}\,L_0$ conspiring with the $\pm i$ in the central charges. This
is consistent with the pure imaginary $L_0, {\bar L}_0$ expectation values in
(\ref{cbar=-c,l0bar-}) and the way the $i$'s conspire in (\ref{KdS3entr+2}).
This is of course not an ab initio explanation of the underlying
spectrum of excitations in such non-unitary CFTs, but a consistent
use of the Cardy formula to view the real-valued Kerr $dS_3$
entropy (\ref{KdS3entr+}), (\ref{KdS3entr+2}).
The complex conformal dimensions $1\pm is$ are akin to those for
operators with dimensions $\Delta_\pm$ dual to bulk scalar modes with 
large mass $\mu$\ (recall recent discussions of the underdamped case 
in \cite{Dey:2024zjx}). Then (\ref{cardy}) gives
\be
\Delta_\pm={d\over 2}\pm\sqrt{{d^2\over 4}-\mu^2l^2}
\quad\xrightarrow{d=2}\quad
S 
\sim \sqrt{ic_0\,(1-is)} + \sqrt{-ic_0\,(1+is)}\ \sim\ 2\sqrt{c_0 s}\,,\ \ 
\ee
for large imaginary parts $s\sim \mu l$. The other imaginary
contributions (suppressed here) are subleading to this real growth 
asymptotically. This gives a picture of de Sitter entropy arising
from dual CFT microstates, albeit only heuristic.

\subsection*{On the Cardy formula for imaginary central charge}

We have employed the Cardy formula (\ref{cardy}) with the CFT data
(\ref{cbar=-c,l0bar-}) comprising imaginary central charges and complex
weights to recover $dS_3$ entropy (\ref{cardydS3ent}) as well as
(\ref{KdS3entr+2}). One might worry that this is not justified for such
exotic non-unitary theories: so it is worth revisiting the derivation
of the Cardy formula.

\medskip

Modular invariance exchanges the spatial and temporal circles and
implies\ $Z[\beta]=Z[{4\pi^2\over\beta}]$.\
In terms of the density of states $\rho(\Delta)$ at weight $L_0=\Delta$, we have the representation\ 
$Z[\beta] = \int\,d\Delta\,\rho(\Delta)\,e^{-\beta (\Delta-{c\over 24})}$,\
which gives\ $Z[\beta]\sim exp[{c\over 24} {4\pi^2\over\beta}]+\ldots$,\
assuming this is dominated by the lowest weights after the modular
transformation. Inverting schematically gives
\be\label{Cardy-Imc}
\rho(\Delta)\sim \int d\beta\,e^{\beta (\Delta-{c\over 24})}\,
\Big(e^{{c\over 24} {4\pi^2\over\beta}}+\ldots\Big)\quad\ra\quad
-{c\pi^2\over 6} {1\over\beta^2}+\Delta=0\,,
\ee
where on the right-hand side, we have evaluated the saddle point value 
that dominates this integral.\ For the case at hand with 
(\ref{cbar=-c,l0bar-}), we have
\be
c=ic_0\,,\ \ \Delta=-i\Delta_0 \quad\ra\quad
\beta=+i\pi\sqrt{c_0\over 6\Delta_0}
\quad\ra\quad \rho(\Delta) \sim e^{2\pi\sqrt{c_0 \Delta_0\over 6}}\,,
\ee
and this choice of $\beta$ makes the $\pm i$'s conspire to give a  leading large real positive asymptotic growth. The saddle value here 
corresponds to imaginary temperature, unconventional but perhaps consistent
with the imaginary weights here. The argument above is of course schematic: among the many questions is the basic point, that defining the CFT partition function relies on a consistent underlying picture of the spectrum of CFT states, which we do not understand here with imaginary central charges and 
complex weights. But the heuristic argument above suggests that perhaps the specific complexifications here do allow a consistent generalization of the 
Cardy formula, and thus might allow a more systematic understanding of the 
dual CFT microstates responsible for de Sitter entropy.

\paragraph{Inner horizon entropy:} This area at $r_-$ can be seen to arise (analogous to (\ref{KdS3entr+2})) as
\bea\label{KdS3entr-}
S_{r_-} = {\pi r_-\over 2G}
&=& {\pi\over 2G} \left({1\over 2}\sqrt{(1-8GE)l^2+8Gjl}
- {1\over 2}\sqrt{(1-8GE)l^2-8Gjl} \right) \nn\\
&=&
2\pi \sqrt{{c \ell_0\over 6}} - 2\pi \sqrt{{{\bar c} {\bar\ell_0}\over 6}}\ ,
\eea
using (\ref{cbar=-c,l0bar-}), and we match the bracket-squared with
$r_-^2$ in (\ref{KdS3r+r-}). This "wrong sign" inner horizon Cardy formula has been previously investigated in \cite{Castro:2012av, Detournay:2012ug}.

\subsection{BTZ black hole entropy: continuations and comparisons}
\label{sec:BTZcontns}

It is interesting to now recall $AdS_3$ BTZ black holes
\cite{Banados:1992wn}\ (in conventions of \cite{Aharony:1999ti})
\bea\label{AdS3BTZ}
ds^2 &=& - {(r^2-r_+^2)(r^2-r_-^2)\over r^2 L^2} dt^2 +
{L^2 r^2\over (r^2-r_+^2)(r^2-r_-^2)} dr^2 +
r^2 \Big(d\phi-{r_+r_-\over r^2 L} dt\Big)^2 \nn\\ [1mm]
&& \qquad 8GM = {r_+^2+r_-^2\over L^2}\,,\qquad  J = {r_+r_-\over 4G L}\,.
\eea
The outer and inner horizons at locations $r_\pm$ here are related 
by $L^2\lra -l^2$ to those in (\ref{KdS3r+r-}).
We now recall Strominger's recovery of the BTZ black hole entropy
via Cardy  \cite{Strominger:1997eq} 
\bea\label{BTZcardy}
&& \qquad\qquad\ L_0+{\bar L_0} = ML\,,\quad  L_0-{\bar L_0} = J\,, \nn\\ [1mm]
&& S = \pi \sqrt{{L (ML+J)\over 2G}} + \pi \sqrt{{L (ML-J)\over 2G}}
= 2\pi \sqrt{{c n_R\over 6}} + 2\pi \sqrt{{c n_L\over 6}}\,,
\eea
with $n_R, n_L$ the $L_0, {\bar L_0}$ eigenvalues: the
holographic values then give the asymptotic growth, viewing the
BTZ-$(M,J)$ black hole for large $M>0$ as excitations over $M=0$.

The complex Kerr $dS_3$ metric (\ref{KdS3:imagJ}) is related by
analytic continuation
\be\label{MJ-Ej-Ll}
L\ra \pm il: \qquad -8GM \equiv 1-8GE\,,\qquad J = \mp ij\,,
\ee
keeping $r_\pm$ fixed, which makes $J$ pure imaginary.
The entropy (\ref{KdS3entr+2}) of the complex Kerr $dS_3$ space is not
quite the analytic continuation from the BTZ entropy (\ref{BTZcardy}):
we have a diagonal CFT constructed from the two CFT copies at
$\scri^\pm$. However, using the mapping (\ref{MJ-Ej-Ll}), it is
instructive to compare complex Kerr $dS_3$ (\ref{KdS3:imagJ}) and
BTZ (\ref{AdS3BTZ}).\ The central charges and weights
(\ref{cbar=-c,l0bar-}) map to $c, L_0, {\bar L_0}$ in (\ref{BTZcardy})
under (\ref{MJ-Ej-Ll}): the diagonal CFT sectors analytically continue
with different signs consistent with the corresponding future/past
boundaries they came from.
Some specific cases are
\bea\label{KdS3-eg}
M=0, J=0\quad &\equiv &\quad 8GE=1,\ j=0\ \ (zero\ entropy);\ \nn\\
M=-{1\over 8G}\quad &\ra&\quad dS_3\ (E=0): \quad
\ell_0 = {1\over 2} \left(-{il\over 8G}\right)\,,\quad
{\bar \ell_0} = {1\over 2} \left({il\over 8G}\right)\,,  \\
J=0 \quad &\ra&\quad {\rm Schw}\ dS_3:\ \ \
\ell_0 = {1\over 2} \left(-{il\over 8G} + ilE\right),\quad
{\bar \ell_0} = {1\over 2} \left({il\over 8G} - ilE\right) ,\nn\\
M\ra large\quad &\Rightarrow&\quad E\gg {1\over 8G}\ \ (disconnected\
from\ usual\ dS\ branch,\  8GE<1)\,.\nn
\eea
So pure $dS_3$ is like the NS vacuum in the discussion of  \cite{Strominger:1997eq},
and this has maximal entropy
$S={\pi l\over 2G}$ and the rest are $(E,j)$-excitations, with a
vanishing entropy at $8GE=1$.  The central point from (\ref{MJ-Ej-Ll})
is that $M<0$ in the $dS$ range, and conversely, large $M$ in the
BTZ black hole does not feature in the conventional $dS$ range.

\medskip

A distinct Kerr $dS_3$ space arises under the analytic continuation
$L\ra il$, keeping $E, J$ fixed. Now $J$ continues to be real, giving
the real Kerr $dS_3$ metric discussed previously in
\cite{Balasubramanian:2001nb} and also 
\cite{Park:1998qk,Klemm:2001ea,Balasubramanian:2002zh,Hao:2026xtk},
\bea\label{KdS3:realJ}
ds^2 = {(r^2-r_+^2)(r^2+r_-^2)\over r^2 l^2} dt^2 -
{l^2 r^2\over (r^2-r_+^2)(r^2+r_-^2)} dr^2 +
r^2 \Big(d\phi-{r_+r_-\over r^2 l} dt\Big)^2\,.
\eea
This real metric has only an outer horizon $r_+$, where the blackening
factor vanishes, with nothing special at $r=r_-$.
By contrast the complex metric (\ref{KdS3:imagJ}) exhibits horizons
at both $r_\pm$, where the blackening factor vanishes, leading to
new features as we have seen. This also leads to interesting
flat space limits, as we will see in what follows.

\section{Ice and Fire}\label{Fire&Ice-0}
The fundamental theories that build up the known universe, including the framework of quantum field theory, are based on the principles of the special theory of relativity. But most of the physical processes perceived in day-to-day life are non-relativistic and governed by Galilean symmetry instead of Lorentz (or Poincar\`e) symmetry. This is the low-energy world, where typical velocities $v$ are much smaller than the speed of light $c$. A route to go from relativistic to non-relativistic symmetries is taking $c\to \infty$. In terms of symmetry algebras, this limit is perceived as an In{\"o}n{\"u}-Wigner contraction of the Poincar\`e algebra to the Galilean algebra.
The contraction of the Poincar\`e algebra (ISO(d,1)) to the Galilean algebra G(d,1) can be realised by scaling the spacetime coordinates $(t,\vec{x})$ as 
\begin{equation}\label{eq: gscale}
    t\to t\,,\quad \vec{x}\to \epsilon\vec{x}\,,\quad \epsilon\to 0\,.
\end{equation}
The Galilean generators obtained after the contraction \eqref{eq: gscale} are,
\begin{align}\label{gal-gen}
H=\partial_t\,,~~P_i=\partial_i\,,~~B_i=t\partial_i\,,~~J_{ij}=-(x_i\partial_j-x_j\partial_i)\,.   
\end{align}
The non-trivial commutation relations of the Galilean generators are given as\footnote{In our notation, $\delta_{k[i}P_{j]}=\delta_{ki}P_j-\delta_{kj}P_i$.}
\begin{equation}\label{eq: gal algebra}
    [J_{ij},J_{kl}]=\text{so}(d)\,,~~ [J_{ij},P_k]=\delta_{k[i}P_{j]}\,,~~ [J_{ij},B_k]=\delta_{k[i}B_{j]}\,,~~[H,B_i]=P_i\,.
\end{equation}
Since the Galilean theories are obtained in the low-energy approximation of relativistic theories, this is the limit we will call ``ice''. 

\medskip

There is a diametrically opposite limit of relativistic theories where, instead of taking the speed of light to infinity, we take $c\to0$. This is the Carrollian limit, where lightcones close up instead of opening out as in the Galilean limit. This leads to ultra-locality. The Carroll algebra in $d$ dimensions is obtained from the $c\to0$ limit of the Poincar\`e algebra. The In{\"o}n{\"u}-Wigner contraction in this case is realised by scaling down the $t$ coordinate instead of the $\vec{x}$ coordinate ($t\to \epsilon t, \, \vec{x}\to\vec{x}\,,\,  \epsilon\to 0$). The Carroll generators obtained after the contraction are,
\begin{align}\label{car-gen}
H=\partial_t\,,~~P_i=\partial_i\,,~~C_i=x_i\partial_t\,,~~J_{ij}=-(x_i\partial_j-x_j\partial_i)\,.   
\end{align}
The non-trivial commutation relations of the Galilean generators are
\begin{equation}\label{eq: car algebra}
    [J_{ij},J_{kl}]=\text{so}(d)\,,~~ [J_{ij},P_k]=\delta_{k[i}P_{j]}\,,~~ [J_{ij},C_k]=\delta_{k[i}C_{j]}\,,~~[P_i,C_j]=\delta_{ij}H\,.
\end{equation}
Carrollian symmetries arise on generic null manifolds and from the point of view of the embedding space can be thought of as an ultra-relativistic limit. To visualise this, think of a theory living in a three dimensional temporal (or spatial) hypersurface of four dimensional Minkowski spacetimes. The 3d theory enjoys $ISO(2,1)$ (or $ISO(3)$)  symmetry. One can use 4d Lorentz boosts to boost this surface. In the limit of infinite boost, the 3d hypersurface becomes null and 3d Carrollian symmetries arise on the hypersurface replacing 3d Poincar\`e symmetry. The Carroll limit is thus an ultra-relativistic boost in this way of understanding. We will call this limit ``fire'', as in contrast to the icy Galilean limit, this would relate to the very high energy sector of the parent relativistic theory. 

\medskip

Before we go further, we note the difference between the two algebras. Notice that in both cases, \cref{eq: gal algebra,eq: car algebra}, the respective boosts commute. But a crucial difference arises in the algebra of the translation and boost generators. In the Galilean case, the algebra of the Hamiltonian and the boosts close onto the spatial momenta. In the Carroll case, the algebra of the spatial momenta and the boosts closes onto the Hamiltonian. In the Carroll case, the Hamiltonian belongs to the centre of the algebra.

\subsection{The curious case of two dimensions}
Rather remarkably, in two spacetime dimensions, the Galilean and the Carrollian algebras are isomorphic to each other. This can be seen by comparing the above vector field representations \eqref{gal-gen} and \eqref{car-gen}. In $d=2$, there are of course no rotation generators. The rest of the algebra is isomorphic. This is easily seen if we exchange the time and space directions, which leads to an identification of $P \leftrightarrow H$ in the two algebras. 

\medskip

The isomorphism extends to their conformal cousins, viz. the Galilean and the Carrollian conformal algebras, in two dimensions. These come from two different contractions of the parent relativistic CFT and the two copies of the Virasoro algebra which dictates its symmetries. Consider the generators of the Witt algebra in 2d CFTs on the Euclidean cylinder. These are given by
\begin{equation}
l_n=\frac{i}{2}e^{in(\tau+\sigma)}(\partial_\tau+\partial_\sigma)\,,\quad \bar l_n= \frac{i}{2}e^{in(\tau-\sigma)}(\partial_\tau-\partial_\sigma)\,.
\end{equation}
The 2d Galilean conformal generators can be obtained by the following combination of the 2d CFT generators, along with the contraction $\sigma\to\epsilon\sigma$ as $\epsilon\to0$ (which is equivalent to sending the speed of light to infinity): 
\begin{equation}\label{eq: GCA}
L_n=l_n+\bar l_n=i e^{i n \tau } \left(\del_\tau+i n \sigma \del_\sigma\right)\,,\quad M_n=\epsilon(l_n-\bar l_n)=ie^{in\tau}\partial_\sigma\,.
\end{equation}
The 2d Galilean conformal algebra (GCA) is then given as
\begin{equation}
[L_n,L_m]=(n-m)L_{n+m}\,,~~[L_n,M_m]=(n-m)M_{n+m}\,,~~[M_n,M_m]=0\,.
\end{equation}
On the other hand, the 2d Carrollian conformal generators can be obtained by the following combination of the 2d CFT generators, along with the contraction $\tau\to\epsilon\tau$ as $\epsilon\to0$ (which sends the speed of light to zero instead of infinity)
\begin{equation}\label{eq: CCA}
L_n=l_n-\bar l_{-n}=i e^{i n \sigma } \left(\del_\sigma+i n \tau \del_\tau\right)\,,\quad M_n=\epsilon(l_n+\bar l_{-n})=ie^{in\sigma}\partial_\tau\,.
\end{equation}
The conformal Carroll algebra (CCA) is then given as
\begin{equation}
[L_n,L_m]=(n-m)L_{n+m}\,,~~[L_n,M_m]=(n-m)M_{n+m}\,,~~[M_n,M_m]=0\,.
\end{equation}
Notice that GCA is isomorphic to the CCA. As expected, the vector fields, \cref{eq: GCA,eq: CCA}, map onto each other when we make the replacement $\tau\leftrightarrow\sigma$. The centrally extended algebra in both cases is given as
\begin{subequations}\label{bms3}
  \begin{align}
& [L_n,L_m]=(n-m)L_{n+m}+c_L(n^3-n)\delta_{n+m,0}\,,\\ 
& [L_n,M_m]=(n-m)M_{n+m}+c_M(n^3-n)\delta_{n+m,0}\,,\\
& [M_n,M_m]= 0. 
\end{align}  
\end{subequations}
The identification of the central charges $c_L\,,\,c_M$ under the contraction from the $\text{Vir} \times \overline{\text{Vir}}$ algebra distinguishes between the Galilean and the Carrollian case. The central charges are identified as
\begin{subequations}\label{iden}
\begin{align}
\text{Galilean CFT:}& \quad c_L=\frac{c+\bar c}{12}\,,\quad c_M=\frac{\epsilon}{12}(c-\bar c)\,, \\
\text{Carrollian CFT:}& \quad c_L=\frac{c-\bar c}{12}\,,\quad c_M=\frac{\epsilon}{12}(c+\bar c)\,.
\end{align} 
\end{subequations}
The above identifications will be a principal diagnostic in our holographic analysis. 

\subsection{Holography}\label{Fire&Ice}
The algebra \eqref{bms3} arises as the asymptotic symmetry algebra of 3d asymptotically flat spacetime at its null boundary and is called the BMS$_3$ algebra \cite{Barnich:2006av}. This identification has led to the formulation of flat holography in terms of co-dimension one non-Lorentzian CFTs \cite{Bagchi:2010zz, Bagchi:2012cy, Bagchi:2016bcd}. Specifically, for Einstein gravity in 3d AFS, the central terms are given by 
\begin{align} \label{flat-cc}
    c_L = 0, \quad c_M = \frac{1}{4G}\,.
\end{align}

\subsection*{Fire -- Flat limit of AdS and Carrollian contraction} 
One of the most obvious routes to flatspace holography is by attempting to understand it as an infinite radius limit of the much better understood AdS examples. Recollect that in AdS$_3$, the asymptotic symmetry algebra is given by two copies of the Virasoro algebra, with two central terms given by 
\begin{align}
    c = \bar{c} = \frac{3 l}{2G}\,,
\end{align}
where $l$ is the radius of AdS. From our identifications above \eqref{iden}, the limit from AdS to flatspace should thus be a Carrollian limit. This is simply because:
\begin{align}
    c_L=\frac{c-\bar c}{12} =0 \,,\quad c_M=\frac{1}{12l}(c+\bar c)= \frac{1}{4G}\,. 
\end{align}
This indeed plays out in the anticipated way. The limit from AdS to flatspace induces a contraction that sends the boundary speed of light to zero and gives a Carrollian CFT.  The formulation of flatspace holography around Carrollian symmetries is called Carrollian holography and has seen a lot of attention in recent times, starting with earlier explorations in 3d bulk and 2d boundary \cite{Bagchi:2010zz, Bagchi:2012cy, Bagchi:2012xr, Bagchi:2012yk, Bagchi:2013lma, Bagchi:2014iea, Jiang:2017ecm, Hartong:2015usd} and then graduating to higher dimensions beginning with \cite{Bagchi:2016bcd, Bagchi:2022emh, Donnay:2022aba}. Recent reviews include \cite{Bagchi:2025vri, Ruzziconi:2026bix, Nguyen:2025zhg}.

\subsection*{Ice -- Flat limit of dS: From the diagonal dual to Galilean CFT}
We have seen that in our construction of the diagonal dual to dS$_3$, we end up with a 2d CFT with two copies of the Virasoro algebra:
\begin{align}
    [\L^\pm_n, \L^\pm_m] = (n-m) \L^\pm_{n+m} + \frac{c^\pm}{12}\delta_{n+m} (n^3-n), \quad [\L^\pm_n, \L^\mp_m] = 0\,. 
\end{align}
Here crucially, as we saw from (\ref{dS3CFT2:Tzz-c,Tzbzb-cbar-1}),
\begin{align}
   c^+ = \frac{3 i l}{2G}, \quad c^- = -\frac{3 i l}{2G}\,.
\end{align}
The combinations that lead to the flat space answers \eqref{flat-cc} are thus
\begin{align}
    c_L=\frac{c^+ + c^-}{12} =0 \,,\quad c_M=\frac{1}{12il}(c^+-c^-)= \frac{1}{4G}\,. 
\end{align}
These, according to our diagnostic \eqref{iden}, are the Galilean combinations and hence the limit from dS$_3$ to 3d Minkowski from the boundary contraction from the relativistic 2d diagonal CFT that we have proposed must be a non-relativistic contraction. In the rest of our paper, we will see how the non-relativistic limit plays a crucial role when we consider the dual of 3d flatspace emerging from dS$_3$. 

\paragraph{Bulk isometries:} We will first show that it is indeed possible to organise bulk isometries of dS$_3$ into a Galilean contraction to get to bulk Mink$_3$ isometries. Of course, this is only concerned with the $n=0,\pm 1$ generators of the Virasoro algebra in the asymptotic expansion and hence is not a proof that the limit has to be Galilean. The objective of the current calculation is to show that a Galilean contraction is possible. 

\medskip

Consider the dS$_3$ metric given by \cref{KdS3:imagJ} with $E=j=0$. The geometry has two distinct regions, namely the static patch $(r<l)$ and the future boundary patch $(r>l)$. We will analyse the isometries of these patches and the corresponding Galilean limit to flat space 

Consider the static patch. The embedding equations for dS$_3$ in here are given by 
\begin{gather}\label{eq: embedding eqns1}
    X^0=\sqrt{l^2-r^2}\sinh{\frac{t}{l}}\,,\quad X^1=\sqrt{l^2-r^2}\cosh{\frac{t}{l}}\,,\quad
    X^2=r\cos\phi\,,\quad X^3=r\sin\phi\,.
\end{gather}
The isometries that preserve the dS$_3$ hypersurface in the embedding picture are the generators of the SO$(1,3)$ group which take the form $M_{AB}=X_A\del_B-X_B\del_A$. These generators can be recast as follows
\begin{gather}\label{eq: SPGen}
        \L^{\pm}_0=\frac{i}{2}(M_{23}\pm i M_{01})\,,\quad  
        \\
        \L^{\pm}_1=\frac{1}{2}(M_{03}\pm M_{13}-i(M_{02}\pm M_{12}))\,,\quad \L^\pm_{-1}=-\frac{1}{2}(M_{03}\mp M_{13}+i(M_{02}\mp M_{12}))\nn\,.
\end{gather}
These closes to two copies of the $\mathfrak{sl}(2,\mathbb R)$ algebra: 
\begin{align}\label{eq: 2 copies SL}
    [\L^+_n,\L^+_m]=(n-m)\L^+_{n+m}\,,\quad [\L^-_n,\L^-_m]=(n-m)\L^-_{n+m}\,,\quad [\L^+_n,\L^-_m]=0\,.
\end{align}

We now recast them into Galilean combinations and take the limit $l\to \infty$:    
\begin{align}
    L_n=\L^+_n+\L^-_n\,,\quad M_n=\frac{1}{il}(\L^+_n-\L^-_n)\,.
\end{align}
This leads to the explicit generators: 
\begin{subequations}
\begin{gather}
L_0=i\partial_\phi\,,\quad L_{\pm1}=-ie^{\pm i\phi}\left( r\partial_t+ t\partial_r\pm\frac{it}{r}\partial_\phi\right)\,,\\
M_0=i\partial_t\,,\quad M_{\pm1}=-e^{\pm i\phi}\left(\pm\partial_r+\frac{i}{r}\partial_\phi\right)\,.
\end{gather}    
\end{subequations}
These close to form the Poincar\`e algebra: 
\begin{align}
    [L_n,L_m]=(n-m)L_{n+m}\,,\quad [L_n,M_m]=(n-m)M_{n+m}\,,\quad [M_n,M_m]=0\,,
\end{align}
Consider now the future boundary patch. The embedding equations here are given as 
\begin{equation}\label{eq: embedding eqns2}
    X^0=\sqrt{r^2-l^2}\cosh{\frac{t}{l}}\,,\quad X^1=\sqrt{r^2-l^2}\sinh{\frac{t}{l}}\,,\quad
    X^2=r\cos\phi\,,\quad X^3=r\sin\phi\,.
\end{equation}
The geometry admits the following Killing vectors in the future boundary patch
\begin{gather}\label{eq: FPGen}
        \L^\pm_0=\frac{i}{2}(M_{23}\pm i M_{01})\,,\\ \L^\pm_1=-\frac{1}{2}(M_{13}\pm M_{03}-i(M_{12}\pm M_{02}))\,,\quad \L^\pm_{-1}=-\frac{1}{2}(M_{13}\mp M_{03}+i(M_{12}\mp M_{02}))\nn\,.
    \end{gather} 
The NR limit of the generators given in \cref{eq: FPGen} leads to the 3d Poincar\`e generators 
\begin{subequations}
    \begin{gather}
        L_0=i\partial_\phi\,,\quad L_{\pm1}=-e^{\pm i\phi}\left(\pm r\partial_t\pm t\partial_r+\frac{it}{r}\partial_\phi\right)\,,\\
        M_0=i\partial_t\,,\quad M_{\pm1}=ie^{\pm i\phi}\left(\partial_r\pm\frac{i}{r}\partial_\phi\right)\,.
    \end{gather}
\end{subequations}
So we see that we can arrange the isometries in the dS$_3 \to$ Mink$_3$ limit in the Galilean combinations. This lends further support to our assertion of the limit being non-relativisitc.

\section{Discovering black holes in 3d flat space}\label{sec:CBH}
Einstein gravity in three dimensions does not have any propagating degrees of freedom and hence the discovery of the BTZ black hole is as such a surprise. It is natural to ponder what happens in 3d AFS. The solutions to Einstein's equations in 3d AFS, with Barnich-Comper\'e boundary conditions \cite{Barnich:2006av}, admit the following general form \cite{Barnich:2012aw}: 
\begin{equation}\label{eq: BGG geometries}
ds^2=\Sigma(\varphi)du^2-2dudr+2\left(\Gamma(\varphi)+\frac{u}{2}\del_\varphi\Sigma(\varphi)\right)dud\varphi+r^2d\varphi^2\,.
\end{equation}
Here $\Sigma(\varphi)$ is the mass aspect ratio and $\Gamma(\varphi)$ is the angular momentum aspect ratio. This leads to a rich class of solutions, with the zero modes describing Flat Space Cosmologies that analogues of the BTZ black holes in AdS$_3$. 

\subsection{Revisiting Flat Space Cosmologies}
In these above general geometries, when the mass and angular momentum aspects are constants, \ie\ specifically 
\begin{align}\label{fscmj}
\Sigma(\varphi)=8GM \quad \text{and} \quad \Gamma(\varphi)=-4GJ\,,
\end{align}
the resulting cosmological spacetimes are called Flat Space Cosmologies (FSCs). More in tune with BTZ coordinates \eqref{AdS3BTZ}, FSCs can be recast into 
\begin{equation}\label{eq: FSC}
ds^2= \frac{\hat r^2}{r^2}\Big(r^2-r_0^2\Big) dt^2 -
{r^2dr^2\over \hat r^2(r^2-r_0^2)} + r^2\left( d\phi -{r_0\hat r\over r^2}dt\right)^2\,,
\end{equation}
where 
\begin{align}\label{rr0}
    \hat r=\sqrt{8GM}, \quad r_0=\sqrt{\frac{2G}{M}}J.
\end{align}
Thus, FSCs admit a horizon $r_0$ which shroud a time-like singularity and the spacetime outside the horizon is a cosmology. 

\medskip

One can attribute thermodynamic quantities to the horizon and specifically, an entropy given by the Bekenstein-Hawking area law 
\begin{equation}
    S=\frac{Area}{4G}=\frac{\pi r_0}{2G}=\frac{\pi J}{\sqrt{2GM}}\,.
\end{equation}
One of the early successes of the flatspace holography programme of recasting the dual in terms of non-Lorentzian CFTs was the discovery of a BMS-Cardy formula \cite{Bagchi:2012xr} that counts the density of states in the dual theory and its matching with bulk entropy. The BMS-Cardy formula reads:  
\begin{align}\label{bms-cardy}
    S=2\pi \left(h_L\sqrt{\frac{c_M}{2h_M}} + c_L \sqrt{\frac{2h_M}{c_M}} \right)\,, 
\end{align}
where apart from the central terms, $h_L$ and $h_M$ are the weights of the states in the dual Carrollian CFT under $L_0$ and $M_0$. One obtains this by constructing a partition function in the dual field theory and appealing to a Carrollian version of modular invariance \cite{Bagchi:2012xr, Bagchi:2013qva, Aggarwal:2025hji}. Particularly important is the S-modular transformations which interchanges the circles on the Carrollian degenerate torus. In the holographic calculations, the field theory weights are identified with the bulk angular momenta and mass (appropriately modified to make the Minkowski vacuum sit at $(h_L, h_M) = (0,0)$): 
\begin{align}
    h_L=J\,,~~h_M=M+\frac{c_M}{2}\,.
\end{align}
Plugging in $c_L=0\,,~~c_M=\frac{1}{4G}$, we see that there is an exact agreement with the boundary and bulk entropies. 

\medskip

It is important to note that the FSC can be geometrically viewed as an orbifold of 3d Minkowski spacetime by a translation and a boosted and is also called a shifted boost orbifold.  It is related to the diagonal, mostly plus Minkowski metric in the $(T,X,Y)$ coordinates by the following coordinate transformation
\begin{equation}
    T={\sqrt{r^2-r_0^2\over \hat r^2}}\cosh{(\hat r\phi)}\,,\quad X={\sqrt{r^2-r_0^2\over \hat r^2}}\sinh{(\hat r\phi)}\,,\quad Y=r_0\phi-\hat r t\,.
\end{equation}
Properties of FSCs have been further investigated in e.g. in \cite{Bagchi:2013lma, Aggarwal:2025hji}. 

\paragraph{The limit from AdS:} A very important part in our subsequent analysis will be played by the limit to 3d AFS. As with the Minkowski vacuum which descends in the $l\to \infty$ limit from the AdS vacuum, the zero mode solutions of the asymptotically AdS spacetimes, viz. the BTZ black holes, map to the FSCs in the limit. 

\medskip

To see this, look explicitly at the metric of the rotating BTZ \eqref{AdS3BTZ}. Remember that the outer and inner horizons are given in terms of the mass $M$ and angular momentum $J$ as
\begin{align}
    r_\pm^2 = 4GM l^2 \left(1\pm\sqrt{1 - \left(\frac{J}{Ml}\right)^2}\right).
\end{align}
Now keeping $(M,J)$ of the original BTZ fixed, let's take the AdS radius to infinity. We see that 
\begin{align}
    r_+ \to \sqrt{8GM} l = \hat r l, \quad r_- \to r_0 = \sqrt{\frac{2G}{M}} J.
\end{align}
This just shows that the rotating BTZ metric reduces to the FSC metric \eqref{eq: FSC} in the $l \to \infty$ limit. In the limit, the outer horizon of the BTZ black hole is sent to infinity, while the inner horizon of the BTZ ($r_-$) stays finite in the limit. In essence, the outer region of the BTZ black hole $r>r_+$ is pushed to infinity and only the region $r_-<r<r_+$ is accessible in the $l\to \infty$ limit. The black hole in AdS thus turns into a cosmology in flatspace. 

\medskip

The entropy captured by the BMS-Cardy formula is actually the remnant of the entropy that can be associated with the inner horizon of the BTZ \cite{Riegler:2014bia, Fareghbal:2014qga} : 
\begin{align}
    S_{inner} = 2\pi \left(\sqrt{{c h\over 6}} - \sqrt{{\bar{c} \bar{h}\over 6}}\,\right) ,
\end{align}
where $h, \bar{h}$ are the usual BTZ weights
\begin{align}
    h = l M + J, \quad \bar{h} = l M - J. 
\end{align}

\subsection{From Kerr dS to complex flat black holes}
We now wish to get back to dS and ask what the solution space looks like as we take the limit to flat spacetime. We start off with the complex Kerr $dS_3$ metric (\ref{KdS3:imagJ}). This gives
\bea\label{dSflatCpxBH}
l\ra\infty:\ &&\ ds^2 = -( 1-8GE ) dt^2 +
{dr^2\over 1-8GE - {16 G^2j^2\over r^2}} + r^2 d\phi^2 + i8Gj dt d\phi \nn\\
&& \qquad = -\frac{\hat r^2}{r^2}\left(r^2-r_0^2\right) dt^2
+ {r^2dr^2\over {\hat r^2(r^2-r_0^2)}}
+ r^2\left(d\phi + i {\hat r r_0\over r^2} dt \right)^2\,,
\eea
where 
\begin{equation}\label{eq: r0 rh}
    r_0={4Gj\over\sqrt{1-8GE}}\,,\quad \hat r^2=1-8GE\,.
\end{equation}
This has effectively pushed the outer horizon $r_+$ out to infinity
while retaining the region containing the inner horizon $r_-$ and the
timelike conical singularity at $r=0$. This spacetime now has no time
dependence. The inner horizon $r_-$ is now an ``event horizon'', where
the blackening factor here vanishes, at $r=r_0$.

\medskip

When $j=0$, there is no horizon and we obtain the region near the
conical singularity $r=0$ of Schwarzschild $dS_3$ (\ref{SdS3:J=0}), \ie\ 
\be\label{dS3flatAsympt}
ds^2 = -\big( 1-8GE  \big) dt^2 + {dr^2\over 1-8GE } + r^2 d\phi^2\,.
\ee
With $j\neq 0$, the singularity at $r=0$ continues to be timelike.
We will call this class of black holes, Complex Black Holes ($\mathbb C$BH). 

The asymptotics at large $r$ are given
by noting that the metric (\ref{dSflatCpxBH}) approach
(\ref{dS3flatAsympt}) which is the metric of a conical defect at $r=0$. So the $dS$ future boundary has disappeared as expected.

The metric (\ref{dSflatCpxBH}) is a
solution to flat space 3d Einstein gravity with $R_{\mu\nu}=0$. 3d flatspace is known to not admit black hole solutions \cite{Ida:2000jh}. But the spacetime (\ref{dSflatCpxBH}) with real $j$
is a complex metric and hence avoids the assumptions of the theorem. 
The ``horizon'' at $r=r_0$ in this complex metric is not on the same
footing as a horizon in real general relativity. 
With real $J$ (\ie\ imaginary $j$), the metric (\ref{dSflatCpxBH})
is real but with no horizon now since $j^2\ra -J^2$.

\medskip

It is worth comparing with the 4d Schwarzschild de Sitter black hole
\cite{Gibbons:1977mu}
\be
ds^2 = -\Big( 1-{2Gm\over r} -{r^2\over l^2} \Big) dt^2 +
{dr^2\over 1-{2Gm\over r} -{r^2\over l^2}} + r^2 d\Omega_2^2\,,
\ee
with the zeros of the blackening factor (for 
${m\over l}\leq {1\over 3\sqrt{3}}$) giving real outer cosmological $r_D$ 
and inner black hole $r_S$ horizons satisfying $0<r_S<r_D<\infty$ and 
a curvature singularity at $r=0$. The flat limit $l\ra\infty$ above 
pushes $r_D$ out to infinity, leaving the flat space Schwarzschild black 
hole with horizon at $2Gm$. In 3d, a complex metric is required 
to realize similar features.

\medskip

The $\mathbb C$BHs solve Einstein equations for zero cosmological constant. These can also be mapped to \eqref{eq: BGG geometries} via the following coordinate transformations with \eqref{eq: r0 rh}
\begin{equation}
    u=t-\frac{r}{\hat r^2}-\frac{r_0}{2\hat r^2}\log \left(\frac{r-r_0}{r+r_0}\right),~~ \varphi =\phi+\frac{i}{2 \hat r}\log \left(\frac{r-r_0}{r+r_0}\right),~~\hat r^2=\lim_{l\to\infty}\frac{r_+^2}{l^2}\,,
\end{equation}
where $r_+$ is the outer horizon radius of Kerr dS$_3$ \eqref{KdS3:imagJ}. The $\mathbb C$BH metric in the $(u,r,\varphi)$ is given as
\begin{equation}
   ds^2= -\hat r^2du^2-2du dr+2ir_0\hat rdud\varphi+r^2d\varphi^2\,.
\end{equation}
The $\mathbb C$BH can be interpreted as the zero mode solution of \eqref{eq: BGG geometries} by identifying 
\begin{equation}\label{cbhmj}
    \Sigma(\varphi)=-\hat r^2\,,~~\Gamma(\varphi)=ir_0\hat{r}\,.
\end{equation}
So we see that the mass aspect has become negative, turning the cosmological spacetime back into a black hole and angular momentum aspect is imaginary. 


\paragraph{Mapping to FSC:} We now connect $\mathbb C$BH explicitly to FSCs. This is simply done by an analytic continuation 
\begin{align}
    \hat r\to i\hat r 
\end{align}
on the $\mathbb C$BH metric. This can also be understood by looking at the generic solution to 3d AFS in Einstein gravity given by the metric \eqref{eq: BGG geometries}, and noting that the mass aspect and angular momentum aspects for the FSC are given by \eqref{fscmj} and \eqref{rr0} and that for the $\mathbb C$BH are given by \eqref{cbhmj}. We note that the analytic continuation on ${\hat r}$ has flipped $r, t$, to time and space, giving the cosmology. 

\medskip

Like FSCs, the flat space complex black hole \eqref{dSflatCpxBH} is related to the diagonal, mostly plus Minkowski metric in the $(T,X,Y)$ coordinates by the following coordinate transformation 
\begin{equation}
    T={i\sqrt{r^2-r_0^2\over \hat r^2}}\cos{(\hat r\phi)}\,,\quad X={\sqrt{r^2-r_0^2\over \hat r^2}}\sin{(\hat r\phi)}\,,\quad Y=r_0\phi+i\hat r t\,.
\end{equation}
It can be shown that \eqref{dSflatCpxBH} is an orbifold of the 3d Minkowski spacetime under the complex vector field
\begin{equation}
    \xi=i\hat r(X\partial_T+T\partial_X)+r_0\partial_Y\,.
\end{equation}

\paragraph{Mapping to a massive spinning particle:} 
The metric (\ref{dSflatCpxBH}), as in the first
line, is related to the spinning particle metric \cite{Deser:1983tn} 
\be
ds^2 = 8GM dt^2 + {r^2\, dr^2\over 16G^2J^2-8GMr^2} + r^2d\phi^2 + 8GJ dt d\phi\,,
\ee
by mapping  $E, j$, to $M, J$, as in (\ref{MJ-Ej-Ll}). This metric can be obtained 
from the metric (\ref{AdS3BTZ}) in the flat limit $L\ra\infty$. Then
$8GM=-1$ \ie\ $E=0$ is flat space with no quotient. With $-1<8GM<0$
or equivalently $0<8GE<1$, we obtain a conical deficit at $r=0$. The
range $M<-1$ corresponds to $8GE>1$ with no obvious significance in
the de Sitter parent geometry.

\medskip

The $\mathbb C$BH can be mapped to a massive spinning particle \cite{Barnich:2012aw} characterised by the following metric
\begin{equation}
    ds^2=-(\alpha dt-Rd\phi)^2+d\rho^2+\alpha^2\rho^2d\phi^2\,,~~\alpha^2=-8GM>0\,,~~R=\frac{4GJ}{\alpha}\,,
\end{equation}
via a different analytic continuation. The coordinate transformation and the subsequent analytic continuation
\begin{equation}
    \rho^2=\frac{r^2-r_0^2}{\hat r^2}\,,\quad r_0\to -ir_0\,,
\end{equation}
transform the $\mathbb C$BH \eqref{dSflatCpxBH} to 
\begin{equation}
    ds^2=-(\hat r dt-r_0d\phi)^2+d\rho^2+\hat r^2\rho^2d\phi^2\,.
\end{equation}
The identification $r_0=R$ and $\hat r=\alpha$ maps the $\mathbb C$BH to the massive spinning particle.

\paragraph{Entropy of $\mathbb C$BH:}
In the flat space limit $l\ra\infty$, the horizon of the Kerr dS$_3$ black hole at $r_+$ is
pushed off to infinity and does not survive, as we have seen in \eqref{dSflatCpxBH}. The horizon at $r_0$ where the blackening factor in (\ref{dSflatCpxBH}) vanishes, can be recognised as arising from the $l\ra\infty$ limit of $r_-$ expanding in (\ref{KdS3r+r-}),
\be
r_-^2\ \xrightarrow{\,l\ra\infty\,}\ 
{16G^2j^2\over 1-8GE} \equiv r_0^2\,,
\ee
and the $dS$ scale $l$ has now disappeared. The area of this
horizon gives the entropy
\be\label{CpxBHent}
S = {2\pi r_0\over 4G} = {2\pi j\over \sqrt{1-8GE}}\,.
\ee
Reality/positivity of this entropy stems from real $j$ (and thus real
$r_-$) in the complex metric (\ref{dSflatCpxBH}). We will see below that this entropy can be
seen to arise in the $l\ra\infty$ limit of the inner horizon entropy expressed in a Cardy-like fashion in (\ref{KdS3entr-}).

\subsection{\texorpdfstring{$\mathbb C$BH}{CBH} entropy via BMS Cardy}
We have proposed that the limit to flatspace from dS$_3$ is a Galilean limit. The $\mathbb C$BH thus should be described in terms of an excited state in the putative dual 2d Galilean CFT. The FSCs have been similarly described in terms of thermal states in a 2d Carrollian CFT.

\paragraph{Galilean route to BMS-Cardy:} The BMS-Cardy formula \eqref{bms-cardy} was derived in \cite{Bagchi:2012xr} in terms of the underlying BMS$_3$ algebra \eqref{bms3} and an intrinsic BMS modular invariance, without recourse to any limit. It was also shown in \cite{Bagchi:2013bga} that both Carrollian and Galilean limits also lead to the BMS-Cardy formula \eqref{bms-cardy}. Since our Galilean limit also deals with unfamiliar weights, let us revisit the analysis leading to the BMS-Cardy formula, now more in tune with our ``diagonal'' dS/CFT story. 

\medskip

From the dual 2d GCFT, the BMS-Cardy formula can be motivated as follows. Consider the partition function $Z_G$ of a 2d GCFT:
\begin{align}
    Z_{G} = \text{Tr} e^{2\pi i \sigma L_0} e^{2\pi i \rho M_0} = \sum d(h_L, h_M) e^{2\pi i \sigma h_L} e^{2\pi i \rho h_M}.
\end{align}
We will consider the theory on a torus and hence the $h_M$ weight will be shifted by the central term $c_M$. (We focus on $c_L=0$ in our analysis here.) Let's call this torus partition function $Z^0$
\begin{align}
    Z^0= \text{Tr} \, e^{2\pi i \sigma L_0} e^{2\pi i \rho (M_0- \frac{c_M}{2})} = e^{-\pi i \rho c_M} Z_{G}.
\end{align}
The 2d GCFT inherits a Galilean modular invariance from its relativistic parent. In particular, the relativistic modular transformation is given by $\tau \mapsto \frac{a \tau + b}{c \tau +d}$ (and similarly for $\bar{\tau}$). The Galilean modular parameters are given by $(\tau,\bar{\tau}) = (\sigma + \epsilon \rho, -\sigma + \epsilon \rho)$. Following the limit $\epsilon \to 0$, we arrive at $(\sigma, \rho) \mapsto (\frac{a \sigma + b}{c \sigma +d}, \frac{\rho}{(c \sigma +d)^2})$. In particular, the S-transformation, which exchanges the two circles of the torus is given by 
\begin{align}
    S: \quad (\sigma, \rho) \mapsto \left(-\frac{1}{\sigma}, \frac{\rho}{\sigma^2}\right)
\end{align}
We demand that the partition function of the theory on the torus be modular invariant. This means under S-transformations: 
\begin{align}
    Z^0 (\sigma, \rho) = Z^0\left(-\frac{1}{\sigma}, \frac{\rho}{\sigma^2}\right) \Rightarrow Z_{G} (\sigma, \rho) 
    = e^{\pi i  c_M\rho (1-\sigma^{-2})} Z_{G}\left(-\frac{1}{\sigma}, \frac{\rho}{\sigma^2}\right). 
\end{align}
Finally, we have the density of states via an inverse Laplace transform: 
\begin{subequations}
   \begin{align}
    & d(h_L, h_M) = \int d\sigma d\rho \, e^{2\pi i g(\sigma, \rho)} \, Z_{G}\left(-\frac{1}{\sigma}, \frac{\rho}{\sigma^2}\right),  \\
& \text{where} \quad    g(\sigma, \rho) = \frac{c_M \rho}{2} \left(1 - \frac{1}{\sigma^2}\right) - h_L \sigma - h_M \rho. 
\end{align} 
\end{subequations}
We will evaluate this integral in the saddle point method. Note that the extrema of the function are at 
\begin{align}
    \sigma_*^2 = \frac{c_M}{c_M - 2 h_M}, \quad \rho_* = \frac{ h_L \sigma_*^3}{c_M}.
\end{align}
Now for our limit from dS$_3$, the weights are unusual. In accordance to this, let us put in 
\begin{align}
    h_L = - i \Delta, \quad h_M = - \xi, \quad \text{with} \quad \Delta, \xi \in \mathbb{R}, \quad \xi>0. 
\end{align}
The value of the function $g$ at the extremum then is given by:
\begin{align}
    g_*= g(\sigma_*, \rho_*) = i \Delta \sigma_*. 
\end{align}
This leads to the entropy (when $\xi \gg c_M$)
\begin{align}\label{bms-cardy2}
    S_G = \log d(\Delta, \xi) = - 2 \pi \Delta \sigma_* = 2\pi |\Delta| \sqrt{\frac{c_M}{2 \xi}}.
\end{align}
Depending on the sign of $\Delta$, the positivity of the entropy picks out one of the two saddles. 

\paragraph{Connecting to $\mathbb C$BH:} 
For our Galilean limit using the diagonal CFT data in (\ref{cbar=-c,l0bar-}), we see that
\bea
&& c_L = {1\over 12} (c+{\bar c}) = 0\,,\quad
h_L = \ell_0+{\bar \ell_0} = -ij\, \Rightarrow \Delta = j.\nn\\
&& c_M = {1\over 12}\,{1\over il} (c-{\bar c}) = {1\over 4G}\,,
h_M = {1\over il} (\ell_0-{\bar \ell_0}) = {8GE-1\over 8G} \leq 0 \Rightarrow \xi = \frac{1-8GE}{8G}
\eea
Plugging into the BMS-Cardy formula \eqref{bms-cardy2} we derived above for these unfamiliar weights, we find the entropy of the 3d $\mathbb C$BH from the dual 2d GCFT:
\begin{align}\label{CBH-Cardy}
   \quad S_{G} = 2\pi |\Delta| \sqrt{\frac{c_M}{2 \xi}} = \frac{2 \pi |j|}{\sqrt{1-8GE}}= S_{{\mathbb C}BH}.
\end{align}

As expected, we can also understand this as a limit of the inner horizon entropy or as a limit from the 2d relativistic "wrong-sign" Cardy formula. To do this, we can recasting as
\be
c=6ilc_M\,,\ \ {\bar c}=-6ilc_M\,,\qquad
\ell_0={1\over 2}(h_L+ilh_M)\,,\ \ {\bar\ell}_0={1\over 2}(h_L-ilh_M)\,,
\ee
we evaluate the entropy combination (\ref{KdS3entr-}) in the flat
space limit $l\ra\infty$ to obtain 
\bea
S_{inner} &=&
 2\pi \left(\sqrt{{c \ell_0\over 6}} - \sqrt{{{\bar c} {\bar\ell_0}\over 6}}\right)
= \sqrt{2}\pi\left(\sqrt{l^2c_M(-h_M)+ilc_Mh_L}-\sqrt{l^2c_M(-h_M)-ilc_Mh_L}
\right)\nn\\
&=& \frac{2 \pi j}{\sqrt{1-8GE}}.
\eea
where we have expanded (noting that $ih_L>0$ and $-h_M>0$ here) to
pick the nonvanishing subleading term. This recovers the complex black hole entropy (\ref{CpxBHent}), with the nontrivial generalization of (\ref{bms-cardy}) to the complex
weights we have here.

\medskip

Of course, the above recovery of the Bekenstein-Hawking area law from the dual field theory entropy comes with the usual caveats one expects when descending to the flat space dual from dS$_3$. As in the $dS_3/CFT_2$ case discussed around (\ref{Cardy-Imc}), it is far from clear how to define states, so the trace over states in the GCFT partition function is only a formal statement at the moment, as is the density of states. In the usual derivation of the BMS-Cardy formula, these states were assumed in either the highest weight or the induced representation of the underlying BMS$_3$ algebra. In our discussions above, it is unclear which representations would play the equivalent role. The Galilean limit usually has an affinity towards the highest weight representations, hence this may be pivotal in our future explorations. 

\medskip

At the level of the uniform density bulk excited macrostate that the
complex black hole (\ref{dSflatCpxBH}) is, the only quantities
characterizing these macrostates are the energy-momentum expectation
values, \ie\ $\ell_0, {\bar\ell_0}$.

\medskip

Let us emphasise again that the flat space limit from de Sitter
is different from that in $AdS$. Let us offer some more insight into this. As discussed after
(\ref{KdS3-eg}), pure $dS_3$ is the maximum entropy macrostate with
$(E,j)$-excitations decreasing the entropy till we hit a vanishing
point $E_{max}$ where the entropy vanishes. This reflects in the
flat space limit as well. We note again that
\be
h_L =\ pure\ imaginary\,,\qquad h_M\leq 0\,,\qquad
\Big[h_M\Big\vert_{E={1\over 8G}}=0\Big] .
\ee
The maximum value $h_M=0$ arises when $E$ is maximal in the parent
$dS_3$: this is when the space closes up, leaving no entropy.
We are thus in an apparently distinct (complexified) branch of the BMS
global algebra characterizing flat space, with complex weights for
$L_0, M_0$, quite unlike what arises from $AdS$ where $h_L, h_M\geq 0$.
The two branches meet at $h_L=0, h_M=0$: this maps to the $(M,J)=0$
BTZ black hole on the $AdS_3$ side.  On the de Sitter side however, it
maps interestingly to not pure $dS_3$, but rather to the maximal black
hole with vanishing entropy (see (\ref{KdS3r+r-})).  This is
reminiscent of the Nariai limit in higher dimensions, but somewhat distinct in 3d.

\section{Conclusions}

We have proposed a diagonal CFT dual to asymptotically $dS_3$ spacetimes, constructed from the holomorphic and antiholomorphic parts of the dual CFTs at the future and past boundaries respectively, in part inspired by an antipodal map between the two CFT copies. The corresponding diagonal holographic energy-momentum tensor expectation values are constructed from the Wavefunction at the future boundary and its conjugate, as in (\ref{TzzPsi-TzbzbPsi*}). This gives equal and opposite imaginary central charges (\ref{dS3CFT2:Tzz-c,Tzbzb-cbar-1}) for the two sectors. In this formulation, pure $dS_3$ as well as the family of complex Kerr $dS_3$ spacetimes with two real horizons correspond to dual CFT data comprising  imaginary central charges and complex weights (\ref{cbar=-c,l0bar-}).
Using this, we recover the entropy of $dS_3$ (\ref{cardydS3ent}) and the complex Kerr family (\ref{KdS3entr+2}) by a Cardy formula.

\medskip

With the central charges as above, the flatspace limit of our dS construction naturally induces a Galilean contraction on the dual diagonal 2d CFT in contrast to the more familiar Carrollian limit when 3d Asymptotically Flat Spacetime (AFS) is approached from $AdS_3$, as we saw in sec.~\ref{Fire&Ice}. The corresponding limit for the complex Kerr family above leaves the inner horizon intact, resulting in flat space complex black holes ($\mathbb{C}$BH) as we saw in (\ref{dSflatCpxBH}): these are closely related to flat space cosmologies which arise from BTZ black holes in AdS$_3$.  We recover the entropy of these $\mathbb{C}$BHs by the BMS-Cardy formula (\ref{CBH-Cardy}). 

\medskip

A noteworthy point is that the Cardy recovery of the entropies, both in the de Sitter case and the flat version, is non-standard given the complex-valued CFT data. What we have isolated are the contributions giving a large real growth, while suppressing other subleading real and imaginary pieces. While several questions arise, this suggests that the underlying spectrum of CFT microstates can be organized in appropriate ways in these sorts of exotic non-unitary CFTs, with a rich structure underneath. We hope to explore these and other issues further in the future.

\bigskip 

\subsection*{Acknowledgements} 
AB and SRI thank the warm hospitality of CMI, where these studies began about a year ago, and KN thanks the warm hospitality of IIT Kanpur as this work developed. AB's research is partially supported by ANRF grants:  ANRF/ARGM/2025/000653/MTR and ANRF/CRG/2022/006165. AB also gratefully acknowledges the support of the Gireesh Jankinath Chair Professorship at IIT Kanpur. SRI thanks Diptarka Das and Sachin Grover for various discussions. SRI is supported by the Institute Assistantship at IIT Kanpur. KN's research is partially supported by a grant to CMI from the Infosys Foundation.

\bigskip

\section*{Appendices}

\appendix

\section{Euclideanization in the static patch\
  \texorpdfstring{$r_-\leq r\leq r_+$}{}:}\label{KdS3Euclst}

The Kerr $dS_3$ metric (\ref{KdS3:imagJ}) in the static patch has
the $t$ coordinate being timelike. Euclideanizing this reveals a
description in terms of a squashed twisted 3-sphere, generalizing
the familiar description of Euclidean pure $dS_3$ as a regular $S^3$.\
To elaborate, we consider the complex metric (\ref{KdS3:imagJ}) and
Wick rotate as\ $t=-it_E$.\ The resulting Euclideanized metric is
real in the static patch\ $r_-\leq r\leq r_+$:
\bea\label{KdS3eucl}
ds^2 &=& {(r_+^2-r^2)(r^2-r_-^2)\over r^2 l^2} dt_E^2 +
{l^2 r^2\over (r_+^2-r^2)(r^2-r_-^2)} dr^2 +
r^2 \Big(d\phi+{r_+r_-\over r^2 l} dt_E\Big)^2 \\
&=& \Big( 1-8GE - {r^2\over l^2} - {16 G^2j^2\over r^2} \Big) dt_E^2
+ {dr^2\over 1-8GE - {r^2\over l^2} - {16 G^2j^2\over r^2}}
+ \Big(r d\phi + {4Gj\over r} dt_E \Big)^2 . \nn
\eea
For $E=0, j=0$, this is the $dS_3$ static patch Euclideanized, which
is $S^3$. With $E,j\neq 0$, the two horizons have distinct periodicities,
wrecking the smooth $S^3$.\ 
To elucidate the structure, it is useful to regard the Lorentzian
Kerr $dS_3$ as a hyperbolic hypersurface in 4-dim Minkowski space,
which we then Euclideanize as
\be
(X^0_E)^2 + (X^1)^2 + (X^2)^2 + (X^3)^2 = l^2\quad \in\quad
ds^2=(dX^0_E)^2+(dX^i)^2 \qquad [X^0=-iX^0_E]\,.
\ee
The global description\
$ds^2=l^2d\tau_E^2+l^2\cos^2\tau_E(d\theta_1^2+\sin^2\theta_1d\theta_2^2)$\
is parametrized as
\be\label{EuclKdS3-global}
X^0_E = l\sin\tau_E\,,\ \
X^1 = l\cos\tau_E\,\cos\theta_1\,,\ \
X^2 = l\cos\tau_E\,\sin\theta_1\cos\theta_2\,,\ \
X^3 = l\cos\tau_E\,\sin\theta_1\sin\theta_2\,,
\ee
while the static coordinatization (\ref{KdS3eucl}) above is given by
\bea\label{EuclKdS3-static}
X^0_E=l\sqrt{\frac{r_+^2-r^2}{r_+^2-r_-^2}}\cos\left(\frac{r_+ }{l^2}t_E+\frac{r_-}{l}\phi\right)\,, &&
X^1=l\sqrt{\frac{r_+^2-r^2}{r_+^2-r_-^2}}\sin\left(\frac{r_+ }{l^2}t_E+\frac{r_-}{l}\phi\right)\,,\nn\\
X^2=l\sqrt{\frac{r^2-r_-^2}{r_+^2-r_-^2}}\cos\left(\frac{r_+}{l}\phi+\frac{r_-}{l^2}t_E\right)\,, &&
X^3=l\sqrt{\frac{r^2-r_-^2}{r_+^2-r_-^2}}\sin\left(\frac{r_+}{l}\phi+\frac{r_-}{l^2}t_E\right)\,.
\eea
In the overlapping patches, matching $X^2, X^3$, and $X^0_E, X^1$,
from (\ref{EuclKdS3-global}) and (\ref{EuclKdS3-static}) gives
\bea
&& r=\sqrt{(r_+^2-r_-^2)\cos^2\tau_E\,\sin^2\theta_1+r_-^2}\,\qquad
\theta_2=\frac{r_+}{l}\phi+\frac{r_-}{l^2}t_E\,, \nn\\
&& \qquad\qquad \frac{r_+ }{l^2}t_E+\frac{r_-}{l}\phi =
\arctan\left({\cos\theta_1\over\tan\tau_E}\right) .
\eea
For $r_+=l,\ r_-=0$, these can be seen to recover the familiar
coordinate relations for the Euclideanized static patch of pure $dS_3$.
In the current case, $\theta_2$ is twisted from $\phi$ by $t_E$ due
to the rotation, and has warped periodicity. The horizon locations
$r=r_\mp$ map to $\theta_1=0, {\pi\over 2}$, on the $\tau_E=0$ slice
(which maps to a nontrivial $(t_E,\phi)$-slice above).\\
Overall, this embedding description shows that the Lorentzian complex
Kerr $dS_3$ metric (\ref{KdS3:imagJ}) arises by a Wick rotation from
the Euclidean $dS_3$ 3-sphere with squashing and twisting encoded by
energy and angular momentum $E,j$, parametrized as above.

\section{Flat space complex black holes: Euclidean action}

Euclideanizing the flat space complex black hole metric
(\ref{dSflatCpxBH}) as\ $t=-it_E$\ gives 
\be
ds^2 = \Big( 1-8GE - {16 G^2j^2\over r^2} \Big) dt_E^2
+ {dr^2\over 1-8GE - {16 G^2j^2\over r^2}}
+ \Big(r d\phi + {4Gj\over r} dt_E \Big)^2 .
\ee
This is a real metric. It can also be obtained as the flat limit
$l\ra\infty$ of the Euclideanized Kerr $dS_3$ metric (\ref{KdS3eucl}).
The time periodicity, \ie\ the inverse temperature $\beta$, is\
(using (\ref{eq: r0 rh}))
\be\label{FScxBH-beta}
\Delta t_E = {8\pi Gj \over (1-8GE)^{3/2}} = {2\pi r_0\over {\hat r}^2}
\equiv \beta\,.
\ee
It also arises from the $l\ra\infty$ limit of the inner horizon
periodicity\ ${2\pi l^2 r_-\over r_+^2-r_-^2}$\ in (\ref{KdS3eucl}).

Consider now an asymptotic boundary at large $r$ for the complex
black hole (\ref{dSflatCpxBH}) as \eg\ in the case of the flat space
4-dim Schwarzschild black hole \cite{Gibbons:1976ue}.
Using $n^r={1\over\sqrt{g_{rr}}}$, the extrinsic curvature on a
large-$r$ hypersurface is
\bea\label{euclFScxBH-K}
&& K = \frac{1}{\sqrt{g}} \del_r(\sqrt{g} n^r)
= {1\over r}\del_r\left(r\sqrt{1-8GE - {16 G^2j^2\over r^2}}\right)
= {1-8GE \over
r \sqrt{1-8GE - {16 G^2j^2\over r^2}}}\,,\nn\\
&& \qquad\qquad\qquad\qquad\
\sqrt{\gamma} = r \sqrt{1-8GE - {16 G^2j^2\over r^2}}\,.
\eea
The Euclidean action has no bulk contribution since $R=0$. We include
a Gibbons-Hawking boundary term but with with a factor half relative
to the conventional definition (since the bulk term vanishes, there is
no inconsistency). We obtain
\be\label{EuclAction-FScxBH}
I_E = -{1\over 16\pi G}\int_{r\ra\infty} d^2x \sqrt{\gamma} K
= - {\Delta t_E\over 4G} (1-8GE) 
\equiv -{\pi r_0\over 4G}\,.
\ee
With $I_E=\beta F$, using (\ref{FScxBH-beta}), the free energy is
recast as
\be
F = -{4\pi^2 T^2\over 8G\Omega_E^2}\,;\qquad
\Omega_E = {d\phi\over dt_E}\Big\vert_{r_0} = -{{\hat r}\over r_0}\,,
\ee
where $\Omega_E$ is the angular velocity at the horizon $r_0$
using (\ref{eq: r0 rh}) Euclideanized.
This gives
\be
S = -\del_TF\Big\vert_{\Omega_E} = {A(r_0)\over 4G}\,,\qquad
j = -\del_{\Omega_E}F\Big\vert_{T}
\ee
The above results are equivalent to analytic continuations
(\ref{eq: FSC}) of those in \cite{Bagchi:2012xr} for the flat
space cosmologies.

\newpage
\bibliographystyle{JHEP.bst}
\bibliography{refsSdSf}

\end{document}